\documentclass[sn-mathphys,Numbered]{sn-jnl}

\usepackage{graphicx}%
\usepackage{multirow}%
\usepackage{amsmath,amssymb,amsfonts}%
\usepackage{amsthm}%
\usepackage{mathrsfs}%
\usepackage[title]{appendix}%
\usepackage{xcolor}%
\usepackage{textcomp}%
\usepackage{manyfoot}%
\usepackage{booktabs}%
\usepackage{algorithm}%
\usepackage{algorithmicx}%
\usepackage{algpseudocode}%
\usepackage{listings}%
\usepackage{gensymb}

\begin{document}

\title[Article Title]{Holonomic swap and controlled-swap gates of neutral atoms via selective Rydberg pumping}


\author[1,2]{\sur{C. F.} \fnm{Sun}}

\author[1]{\sur{X. Y.} \fnm{Chen}}

\author[3]{\sur{W. L.} \fnm{Mu}}

\author[1,2]{\sur{G. C.} \fnm{Wang}}

\author[4]{\sur{J. B.} \fnm{You}}

\author*[1,2]{\sur{X. Q.} \fnm{Shao}}\email{shaoxq644@nenu.edu.cn}

\affil[1]{\orgdiv{Center for Quantum Sciences and School of Physics}, \orgname{ Northeast Normal University}, \orgaddress{\city{Changchun}, \postcode{130024}, \country{China}}}

\affil[2]{\orgdiv{Key Laboratory for UV Light-Emitting Materials and Technology of Ministry of Education}, \orgname{ Northeast Normal University}, \orgaddress{ \city{Changchun}, \postcode{130024}, \country{China}}}

\affil[3]{\orgdiv{Department of Physics}, \orgname{Beijing Normal University}, \orgaddress{ \city{Beijing}, \postcode{100875}, \country{China}}}

\affil[4]{\orgdiv{Institute of High-Performance Computing}, \orgname{A*STAR (Agency for Science, Technology and Research)}, \orgaddress{\street{1 Fusionopolis Way}, \city{Connexis}, \postcode{138632}, \country{ Singapore}}}

\abstract{Holonomic quantum computing offers a promising paradigm for quantum computation due to its error resistance and the ability to perform universal quantum computations. Here, we propose a scheme for the rapid implementation of a holonomic swap gate in neutral atomic systems, based on the selective Rydberg pumping mechanism. By employing   time-dependent soft control, we effectively mitigate the impact of off-resonant terms even at higher driving intensities compared to time-independent driving. This approach accelerates the synthesis of logic gates and passively reduces the decoherence effects. Furthermore, by introducing an additional atom and applying the appropriate driving field, our scheme can be directly extended to implement a three-qubit controlled-swap gate. This advancement makes it a valuable tool for quantum state preparation, quantum switches, and a variational quantum algorithm in neutral atom systems.}

\keywords{Neutral atom, Holonomic quantum computing, Selective Rydberg pumping}

\maketitle

\section{Introduction}\label{sec1}

Quantum computers, which use the best algorithms currently known, offer the possibility of
solving certain computational tasks much more effectively than any classical
counterpart \cite{Feynman,shor,Freedman,Childs,Hallgren}. This has inspired a great deal of searches for building scalable and functional quantum computers over the past two decades.
Realizing a universal set of quantum gates with high
fidelities is the key to implementing quantum computation.
However, errors in the control process of a quantum system inevitably affect the quantum gate, and the
propagation of these inaccurate control errors may quickly spoil the
practical realization.
Due to the fault tolerance for some types of errors in the control
process, holonomic quantum computation (HQC),  first proposed by Zanardi and Rasetti \cite{Zanardi}, is one of the well-known strategies
to improve gate robustness by using non-abelian geometric phases \cite{Wilczek}.

Early HQC schemes were based on adiabatic evolution \cite{Zanardi,Pachos,Duan}. In this case, one obstacle is the long running time required for adiabatic evolution, which
makes the gates vulnerable to open system effects and
parameter fluctuations that may hinder the experimental implementation.
To overcome this shortcoming, nonadiabatic HQC based on nonadiabatic non-abelian
geometric phases \cite{Anandan} was proposed \cite{Svist}, which has become a promising quantum computation paradigm and has attracted increasing attention recently \cite{Xu,Mousolou,Xu3,Xu1,Herterich,Hong,Zhang,Xu2,Liu,Chen,Zhao1,Wang,Liu1,Shen,Li1}.
Meanwhile, nonadiabatic HQC has been
experimentally demonstrated on various physical platforms,
such as superconducting transmons \cite{Abdumalikov,XuY,YanT}, the nuclear magnetic
resonance system \cite{Feng1,LiH,ZhuZ}, and the diamond nitrogen vacancy center \cite{ZuC,Arroyo,Sekiguchi,ZhouB,Nagata}.

As highly promising candidates for quantum computation and simulation, Rydberg atoms exhibit remarkable attributes that make them particularly attractive for these applications. 
One of their key advantages is their long coherence time for ground-state atoms, combined with the exceptional properties of highly excited Rydberg states. 
These highly excited Rydberg states not only possess an extended lifetime proportional to the third power of the principal quantum number, but also significantly interact through strong long-range Rydberg-Rydberg interactions, manifesting as Rydberg-mediated dipole-dipole or van der Waals interactions \cite{Saffman,Browaeys,Saffman1,ShaoXQ1,10.1063/5.0192602}. 
The presence of these strong Rydberg-Rydberg interactions enables a phenomenon known as Rydberg blockade, where the resonant excitation of two or more atoms to the Rydberg states is hindered \cite{Jaksch,Lukin}. 
Rydberg blockade has been proven instrumental in the effective implementation of quantum logic gates, as experimentally demonstrated with individual atom \cite{Urban,Gaetan}.
Another representative phenomenon observed in neutral atom systems is the Rydberg antiblockade, allowing for a resonant two-photon transition with the energy shift of the Rydberg pair states compensated for by the two-photon detuning \cite{Ates,Amthor,SU2016,SU2017,SU2018,SU202001,Su_202009}. 
Taking advantage of the unique characteristics of Rydberg atoms, they are exceptionally well suited for qubit encoding and serve as an excellent medium in the field of quantum computing and quantum simulation \cite{Mller,Carr,Tian,Su,Shao,Zeng,Shi,Petrosyan,LDX,Omran,Wintermantel,Su1,Bai,Yin:20,Yin:21,Wu:21,SHI2021007,SHI202110,PhysRevApplied.18.044042,Shi_2022,Jandura2022timeoptimaltwothree,Evered2023,SHI202302,Shi202310,PhysRevA.109.042604}.

In contrast to conventional Rydberg blockade or Rydberg antiblockade, the selective Rydberg pumping (SRP) mechanism provides a novel approach to selectively exciting the target quantum states of neutral atoms while effectively freezing the evolution of nontarget quantum states \cite{ShaoXQ}. 
The SRP mechanism capitalizes on the synergistic effects of multifrequency driving fields and strong dipole-dipole interactions, wherein the first-order Rabi coupling supersedes the traditional second-order dynamics required by Rydberg antiblockade. 
The advantage of adopting first-order Rabi coupling lies in its typically much larger magnitude compared to the second-order interaction. Consequently, the SRP-based scheme enables significant acceleration, while concurrently reducing the attenuation of other excited Rydberg states.

In this work, we aim to implement a holonomic swap and controlled-swap gates in neutral atomic systems, using the SRP mechanism \cite{ShaoXQ}. 
However, the previous SRP mechanism featured a time-independent interaction strength. The coupling coefficient, which governs the system's time evolution, must be sufficiently small to achieve improved addressing of resonant terms and efficient suppression of off-resonant interactions.
To overcome these challenges, we propose employing Gaussian time-dependent soft control in the SRP mechanism \cite{Haase,Han1}. This approach effectively mitigates the impact of off-resonant terms even at higher driving intensities compared to time-independent driving. As a result, the synthesis of logic gates is accelerated, and the decoherence effects are greatly reduced.

\section{Holonomic two-qubit swap gate}\label{sec2}
\begin{figure}
\centering
\scalebox{0.6}{\includegraphics{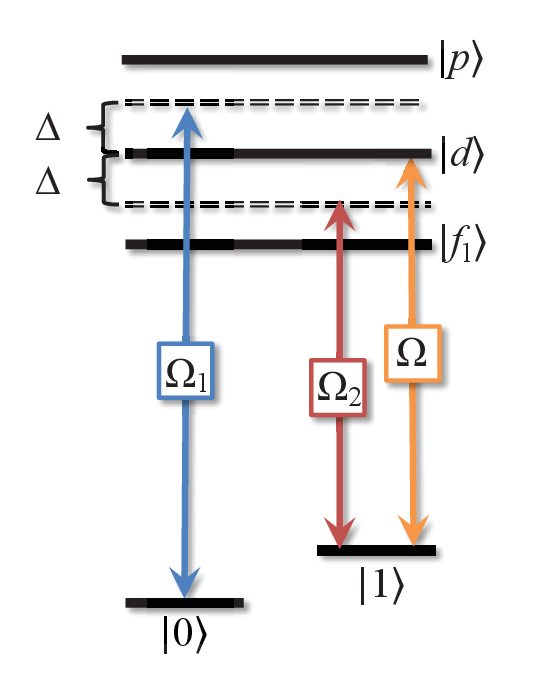}} \caption{\label{p0}
 Schematic view of the atomic-level configuration. Three types of laser fields are applied to drive each atom. One laser field with Rabi frequency $\Omega_{1}$ is applied to drive the transition $|0\rangle \leftrightarrow |d\rangle$
with a blue detuning $\Delta$.
At the same time, the ground state $|1\rangle$ is dispersively coupled to the excited state $|d\rangle$  by one laser field
with Rabi frequency $\Omega_{2}$ (which has a red detuning $\Delta$) and another
resonant laser field with Rabi frequency $\Omega$, simultaneously.
}
\end{figure}
Here we consider a system of two identical $^{87}$Rb atoms, and each atom consists of three long-lived Rydberg states 
$|p\rangle=|61 P_{1/2},m_{J}=1/2\rangle$, $|d\rangle=|59 D_{3/2},m_{J}=3/2\rangle$, and $|f_1\rangle=|57 F_{5/2},m_{J}=5/2\rangle$, and two ground states $|0\rangle=|5 S_{1/2}  F=1,m_{F}=1\rangle$ and $|1\rangle=|5S_{1/2},F=2,m_{F}=2\rangle$. A laser field with Rabi frequency $\Omega_{1}$ is applied to drive the transition $|0\rangle \leftrightarrow |d\rangle$
with a blue detuning $\Delta$ on the order of several hundred megahertz.  At the same time, the ground state $|1\rangle$ is dispersively coupled to the excited state $|d\rangle$ by one laser field
with Rabi frequency $\Omega_{2}$ (which has a red detuning $\Delta$) and another
resonant laser field with Rabi frequency $\Omega$, simultaneously. As demonstrated in Refs. \cite{Sylvain,PhysRevA.92.020701,Browaeys,ashkarin2023highfidelity}, the pair states $|dd\rangle$, $|pf_1\rangle$, and $|f_1p\rangle$ exhibit nearly degenerate characteristics. Consequently, the resonant dipole-dipole (F\"{o}rster resonance) interaction between the two Rydberg atoms causes a hopping transition between the Rydberg states $|dd\rangle$ and $(|pf_1\rangle+|f_1p\rangle)/\sqrt{2}$, with a coupling strength of $\sqrt{2}J$. The Hamiltonian of the two-atom system in the interaction picture reads ($\hbar = 1$)
\begin{equation}\label{initial2}
H_{\mathrm{full}} =\sum_{k=1}^{2}\big[\Omega_{1}e^{i\Delta t}|0\rangle_{k}\langle d|+(\Omega+\Omega_{2}e^{-i\Delta t})|1\rangle_{k}\langle d|\big]+J|dd\rangle(\langle pf_1|+\langle f_1p|)+\mathrm{H.c.},
\end{equation}
where $J=C_{3}/R^{3}$ with $C_{3}/(2\pi)=\mathrm{2.54~GHz~\mu m^{3}}$ and $R$ represents the distance between the atoms.
Due to the strong dipole-dipole interaction between the two Rydberg atoms,  performing a
rotation with the frame defined by 
$\exp[-\sqrt{2}iJt(|E_{+}\rangle\langle E_{+}|-|E_{-}\rangle\langle E_{-}|)]$, where  $|E_{\pm}\rangle=[\sqrt{2}|dd\rangle\pm(|pf_1\rangle+|f_1p\rangle)]/2$
are the eigenstates of the Rydberg dipole-dipole interaction with the eigenvalues $\pm \sqrt{2} J$, respectively, we obtain the
following transformed Hamiltonian  
\begin{eqnarray}\label{initial22}
H_{{\rm full}}&=&H_{1}+H_{2},\\
H_{1}&=&\sqrt{2}\Omega_{S}|00\rangle\langle T_{0}|e^{i\Delta t}+
\frac{1}{\sqrt{2}}|01\rangle\big[(\langle T_{0}|-\langle S_{0}|)(\Omega+\Omega_{S}e^{-i\Delta t})+\langle d1|\sqrt{2}\Omega_{S}e^{i\Delta t}\big]\nonumber\\
&&+\frac{1}{\sqrt{2}}|10\rangle\big[(\langle T_{0}|+\langle S_{0}|)(\Omega+\Omega_{S}e^{-i\Delta t})+\langle 1d|\sqrt{2}\Omega_{S}e^{i\Delta t}\big]\nonumber\\
&&+\sqrt{2}(\Omega+\Omega_{S}e^{-i\Delta t})|11\rangle\langle T_{1}|+\mathrm{H.c.} ,\nonumber\\
H_{2}&=&\Omega_{S}|T_{0}\rangle\big[\langle E_{+}|e^{i(\Delta-\sqrt{2}J)t}+\langle E_{-}|e^{i(\Delta+\sqrt{2}J)t}\big]+|T_{1}\rangle\big[\langle E_{+}|(\Omega e^{-i\sqrt{2}J t}\nonumber\\
&&+\Omega_{S}e^{-i(\Delta+\sqrt{2}J) t})+\langle E_{-}|(\Omega e^{i\sqrt{2}J t}+\Omega_{S} e^{-i(\Delta-\sqrt{2}J) t})\big]+\mathrm{H.c.},\nonumber
\end{eqnarray}
where we have assumed all the Rabi frequencies are real and set $\Omega_{1}=\Omega_{2}=\Omega_{S}$ for simplicity.  $|T_{0}(S_{0})\rangle=(|d0\rangle\pm|0d\rangle)/\sqrt{2}$ and $|T_{1}\rangle=(|d1\rangle+|1d\rangle)/\sqrt{2}$. $H_{1}$ 
describes the transitions between the ground states
and the Rydberg states with a single excitation, and $H_{2}$ bridges the interaction between the Rydberg states with a single excitation and two excitations.
Here we consider the large detuning case, such as $\Delta\gg \{\Omega_{S},\Omega\}$ and $\Delta=\sqrt{2}J$, in this case the terms oscillating with high frequencies
$\{\pm(\Delta+\sqrt{2}J),\pm\Delta,\pm\sqrt{2}J\}$ in Eq.~(\ref{initial22}) can be safely disregarded (see Appendix A for details), then the Hamiltonian can be evaluated explicitly

\begin{eqnarray}\label{initial220}
H &=&\frac{1}{\sqrt{2}}\Omega(|10\rangle-|01\rangle)\langle S_{0}|+\frac{1}{\sqrt{2}}\Omega(|10\rangle+|01\rangle)\langle T_{0}|+\sqrt{2}\Omega|11\rangle\langle T_{1}|+\Omega_{S}|T_{0}\rangle\langle E_{+}|\nonumber\\
&&+\Omega_{S}|T_{1}\rangle\langle E_{-}|+\mathrm{H.c.}.
\end{eqnarray}

{In Ref.~\cite{Haase}  soft temporal quantum
control, which enables on-resonant coupling within a desired
set of target systems and efficiently avoids unwanted off-resonant contributions coming from others, has been proposed. 
Here, we use the technique of soft quantum control and choose the Rabi frequency $\Omega$ as a time-dependent
Gaussian form, $\Omega(t)=\Omega_{m}\mathrm{exp}[-(t-2T)^{2}/T^{2}]$, where
$\Omega_{m}$ and $T$ are the maximum amplitude and width of the
Gaussian pulse, respectively.  The Hamiltonian $H$ in Eq.~(\ref{initial220}) can be rewritten as  
\begin{equation}\label{initial222}
H =H_{\mathrm{S}}+H_{\mathrm{int}},
\end{equation}
where
\begin{eqnarray}\label{HS}
H_{\mathrm{S}} &=&\Omega_{S}|T_{0}\rangle\langle E_{+}|+\Omega_{S}|T_{1}\rangle\langle E_{-}|+\mathrm{H.c.},\nonumber\\
H_{\mathrm{int}} &=&\frac{1}{\sqrt{2}}\Omega(t)(|10\rangle-|01\rangle)\langle S_{0}|+\frac{1}{\sqrt{2}}\Omega(t)(|10\rangle+|01\rangle)\langle T_{0}|+\sqrt{2}\Omega(t)|11\rangle\langle T_{1}|+\mathrm{H.c.}.\nonumber
\end{eqnarray}
Under the eigenvalues $\omega_{j}$ and the corresponding projection operators  $\mathbb{P}(\omega_{j})$ of $H_{\mathrm{S}}$, $H_{\mathrm{S}}$ is reformulated in the diagonal form as $H_{\mathrm{S}}^{1}=\sum_{j}\omega_{j}\mathbb{P}(\omega_{j})$.
 Meanwhile, $H_{\mathrm{int}}$  under the projection operators $\mathbb{P}(\omega_{j})$ is \begin{equation}\label{soft1}
H_{\mathrm{int}}^{1}=\sum_{j,k}
\mathbb{P}(\omega_{j})H_{\mathrm{int}}\mathbb{P}(\omega_{k}).
  \end{equation}
 Our purpose is to suppress the terms with energy mismatches in
Eq.~(\ref{soft1}) for which $\omega_{j}\neq\omega_{k}$, and to keep the energy conserving
ones  for which $\omega_{j}=\omega_{k}$ by adopting the time-dependent control.

 We first analyze the propagator $U_{D}=\mathrm{exp}(-i\int_{0}^{4T}H_{D}dt)$, where $H_{D}=H_{{\rm S}}^{1}+\sum_{j}\mathbb{P}(\omega_{j})H_{\mathrm{int}}\mathbb{P}(\omega_{j})$ includes the desired resonance interactions. It is easy to verify that in the latter all $\mathbb{P}(\omega_{j})H_{\mathrm{int}}\mathbb{P}(\omega_{j})$ operators commute with
$H_{{\rm S}}^{1}$, so
$H_{D}$ can be diagonalized in the common
eigenstates $|\psi^{D}_{j}\rangle$
of $H_{{\rm S}}^{1}$ and $\mathbb{P}(\omega_{j})H_{\mathrm{int}}\mathbb{P}(\omega_{j})$. Therefore $U_{D}= \sum_{j} e^{-i\phi^{D}_{j}(4T)}|\psi_{j}^{D} \rangle \langle \psi_{j}^{D} |$
 is also diagonal in the basis ${|\psi_{j}^{D} \rangle}$
and the dynamic phases
$\phi^{D}_{j}(4T)$ include the effect of energy
shifts from $\mathbb{P}(\omega_{j})H_{\mathrm{int}}\mathbb{P}(\omega_{j})$.

If the whole Hamiltonian   in Eq.~(\ref{initial222}) under the projection operators  $\mathbb{P}(\omega_{j})$, $H^{1}=H_{{\rm S}}^{1}+H_{\mathrm{int}}^{1}$, 
is considered, the time
evolution operator $U=\mathbf{T}\mathrm{exp}(-i\int_{0}^{4T}H^{1}dt)$ with $\mathbf{T}$ being time ordering
is generally non-diagonal in the basis ${|\psi_{j}^{D} \rangle}$ and the non-commuting 
terms $\mathbb{P}(\omega_{j})H_{\mathrm{int}}\mathbb{P}(\omega_{k})$ ($j\neq k$) would cause unwanted transitions between the different states 
$|\psi_{j}^{D} \rangle$.

However, when the soft control is included one can efficiently eliminate the unwanted interactions caused by $\mathbb{P}(\omega_{j})H_{\mathrm{int}}\mathbb{P}(\omega_{k})$ ($j\neq k$)
even for long evolution times.
At the boundaries of the
interaction times ($0$ and $4T$), $\Omega(t)$ has negligible values,
and therefore the eigenstates of the whole system coincide with those of
$H_{D}$.
 More precisely, under the condition of adiabatic evolution \cite{Wangzy,Xuk}, 
there are no transitions among the states $|\psi_{j}^{D} \rangle$ , and the propagator at the end of the evolution is
\begin{equation}\label{evolution}
  U\approx \sum_{j} e^{-i\phi_{j}(4T)}|\psi_{j}^{D} \rangle \langle \psi_{j}^{D} |\equiv \bar{U}\equiv e^{-4i\bar{H}T},
  \end{equation}
where $\phi_{j}(4T)$ are the dynamic phases of $H^{1}$,
while the geometric phases vanish because $\Omega(t)$ returns to its original value \cite{Griffiths}.  In this manner $U$ takes the
same form as $U_{D}$ and the adiabatic average Hamiltonian for
the soft quantum control scheme is
\begin{equation}\label{average1}
  \bar{H}=\sum_{n} \frac{\phi_{j}(4T)}{4T} |\psi_{j}^{D} \rangle \langle \psi_{j}^{D} |.
  \end{equation}
Using this method, 
the Hamiltonian of our current model reduces to an average form (see Appendix B for details)
\begin{eqnarray}\label{average}
  \bar{H}&=&\frac{g}{\sqrt{2}}(|10\rangle-|01\rangle)\langle S_{0}|+\frac{g}{\sqrt{2}}|S_{0}\rangle(\langle 10|-\langle 01|)\nonumber\\
  &&
  +\frac{1}{4T}\int_{0}^{4T}\sqrt{\Omega(t)^{2}+\Omega_{S}^{2}} dt ( |\psi^{D}_{1}\rangle\langle \psi^{D}_{1}|-|\psi^{D}_{2}\rangle\langle \psi^{D}_{2}|)
  \nonumber\\
  &&+\frac{1}{4T}\int_{0}^{4T}\sqrt{2\Omega(t)^{2}+\Omega_{S}^{2}} dt (|\psi^{D}_{3}\rangle\langle \psi^{D}_{3}|- |\psi^{D}_{4}\rangle\langle\psi^{D}_{4}|),
\end{eqnarray}
where $g=\sqrt{\pi}\Omega_{m}\mathrm{Erf}[2]/4$. $|\psi^{D}_{1}\rangle=(|T_{0}\rangle+|E_{+}\rangle)/\sqrt{2}$, $|\psi^{D}_{2}\rangle=(|T_{0}\rangle-|E_{+}\rangle)/\sqrt{2}$, 
$|\psi^{D}_{3}\rangle=(|T_{1}\rangle+|E_{-}\rangle)/\sqrt{2}$, and $|\psi^{D}_{4}\rangle=(|T_{1}\rangle-|E_{-}\rangle)/\sqrt{2}$ are the eigenstates of $H_{\mathrm{S}}$ governed by $\Omega_{S}$.}

In the following, the associated propagator $U=e^{-4i\bar{H}T}$ with the
evolution period $\tau=4T$ can be used to generate a high-fidelity two-qubit swap gate. The time evolution operator $U$ in the basis $\{|S_{0}\rangle,|00\rangle,|01\rangle,|10\rangle,|11\rangle\}$ reads 
 \begin{eqnarray}
U(\tau)=\left(\begin{array}{c c c c c}
\cos\lambda_{\tau} & 0 & \frac{\sqrt{2}i}{2}\sin\lambda_{\tau} &-\frac{\sqrt{2}i}{2}\sin\lambda_{\tau} & 0 \\
0& 1& 0& 0& 0 \\
\frac{\sqrt{2}i}{2}\sin\lambda_{\tau} &0 & \frac{1}{2}(1+\cos\lambda_{\tau}) &\frac{1}{2}(1-\cos\lambda_{\tau}) & 0 \\
-\frac{\sqrt{2}i}{2}\sin\lambda_{\tau} &0 & \frac{1}{2}(1-\cos\lambda_{\tau}) & \frac{1}{2}(1+\cos\lambda_{\tau}) &0 \\
0 & 0 &0 &0 & 1 \\
\end{array}
\right),\nonumber
\end{eqnarray}
where $\lambda_{\tau}=\tau g=4Tg$.
Choosing the parameters to satisfy that
\begin{equation}
\lambda_{\tau}=4Tg=\pi,
\end{equation}
one can derive $T=\sqrt{\pi}/(\mathrm{Erf}[2]\Omega_{m})$. The final effective  evolution operator $U(\tau)$ is
 \begin{eqnarray}
U(\tau)=\left(\begin{array}{c c c c c}
-1 & 0 & 0 &0 & 0 \\
0& 1& 0& 0& 0 \\
0 &0 & 0 &1 & 0 \\
0 &0 & 1 & 0 &0 \\
0 & 0 &0 &0 & 1 \\
\end{array}
\right),
\end{eqnarray}
which is a  two-qubit swap gate on the computational
subspace  ${\mathbf S}=\mathrm{Span}\{|00\rangle,|01\rangle,|10\rangle,|11\rangle\}$ as follows
 \begin{eqnarray}
U_{\mathrm{swap}}=\left(\begin{array}{c c c c}
 1& 0& 0& 0 \\
0 & 0 &1 & 0 \\
0 & 1 & 0 &0 \\
 0 &0 &0 & 1 \\
\end{array}
\right).
\end{eqnarray}

Next, we confirm that the effect of $U(\tau)$ on ${\mathbf S}$ is entirely holonomic.
First, we briefly review the conditions of
nonadiabatic HQC proposed in Refs. \cite{Svist,Xu}. Consider a $N$-dimensional quantum system with a Hamiltonian $H(t)$. Assume that there exists a time-dependent $K$-dimensional subspace $\mathbf K(t)$  spanned by a set of orthonormal bases
$\{|\Phi_{m}(t)\rangle,m=1,\ldots,K\}$ at each time $t$. Here, $|\Phi_{m}(t)\rangle$ can be obtained by the Schr$\mathrm{\ddot{o}}$dinger equation
\begin{equation}
|\Phi_{m}(t)\rangle=\mathbf{T}\exp[-i\int_{0}^{t}H(t')dt']|\Phi_{m}(0)\rangle=U(t)|\Phi_{m}(0)\rangle, \nonumber
\end{equation}
with $\mathbf{T}$ being timing ordering, and $m=1,\ldots,K$. The unitary transformation $U(\tau)=\mathbf{T}\exp[-i\int_{0}^{\tau}H(t')dt']$  is a holonomy matrix acting on the $K$-dimensional subspace $ \mathbf K (0)$ spanned by $\{|\Phi_{m}(0)\rangle\}_{m=1}^{K}$ if $|\Phi_{m}(t)\rangle$ satisfies the following two conditions:
\begin{equation}\label{condition}
\begin{aligned}
(&{\rm I})~~\sum_{m=1}^{K}|\Phi_{m}(\tau)\rangle\langle\Phi_{m}(\tau)|=\sum_{m=1}^{K}|\Phi_{m}(0)\rangle\langle\Phi_{m}(0)|,\\
(&{\rm II})~~\langle\Phi_{m}(t)|H(t)|\Phi_{n}(t)\rangle=0, ~~~ m,n=1,\ldots,K.
\end{aligned}
\end{equation}
 Condition $(\rm I)$ ensures that the states in the subspace
complete a cyclic evolution, and condition $(\rm II)$ ensures that the cyclic evolution is purely geometric.

\begin{figure}
\centering
\includegraphics[width=1\linewidth]{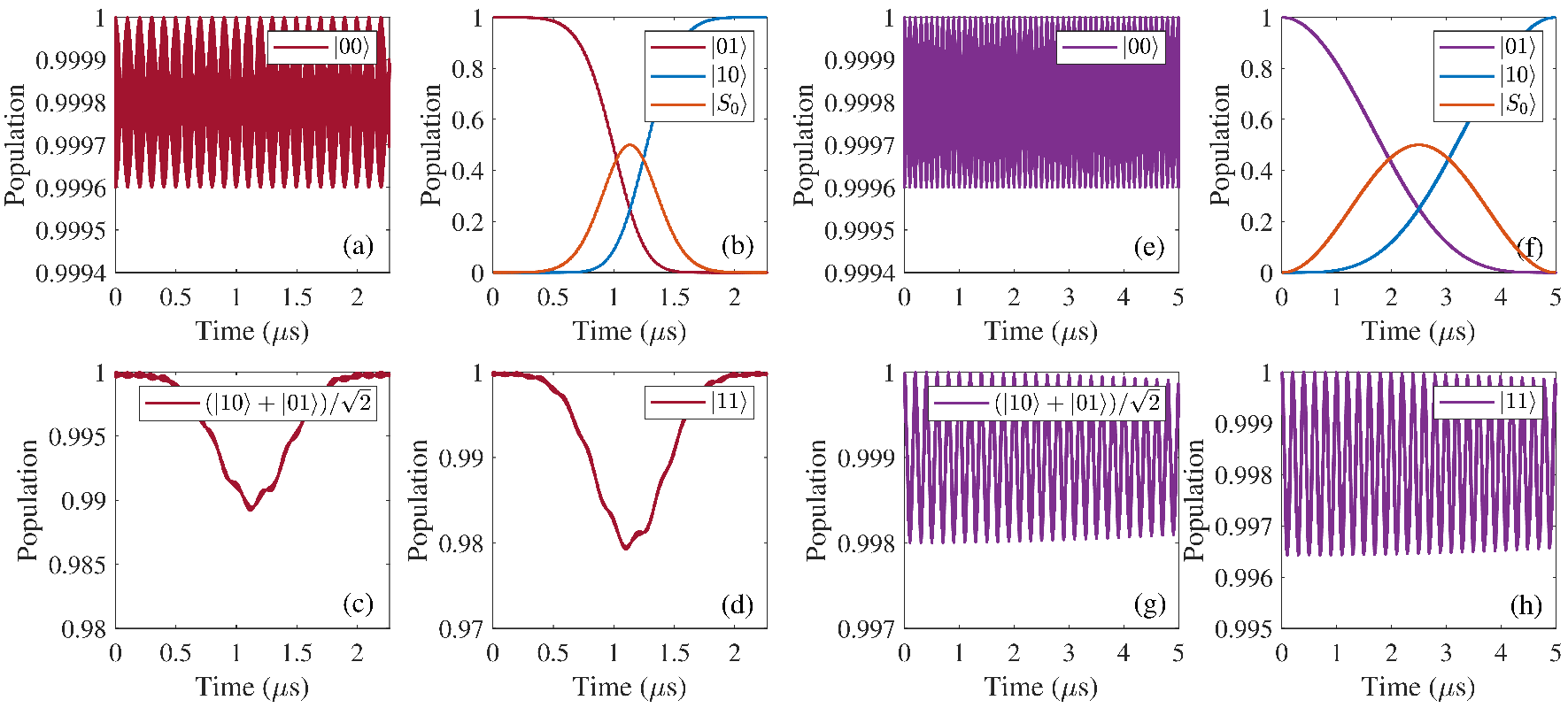} 
\caption{\label{p22}
  The temporal evolution of populations for different ground states $\{|00\rangle,|01\rangle,|10\rangle,|11\rangle\}$ corresponding to (a)-(d) governed by the full Hamiltonian Eq.~(\ref{initial2}). (Left)
The  time-dependent control parameter is $\Omega(t)=\Omega_{m}\mathrm{exp}[-(t-2T)^{2}/T^{2}]$ with $\Omega_{m}/(2\pi)=0.5$ MHz and
$T=\sqrt{\pi}/(\mathrm{Erf}[2]\Omega_{m})$, where the average value of $\Omega(t)$ is $\overline{\Omega(t)}/(2\pi) =0.22$~MHz. 
 (Right)  The corresponding numerical simulations (e)-(f) under the time-independent driving
 $\Omega/(2\pi) =0.1$ MHz. 
For both cases,  the other parameters are the same as $\Omega_{S} /(2\pi) =5$ MHz, 
 $\Delta /(2\pi) =500 \sqrt{2}$ MHz, and  $J/(2\pi)= 500$ MHz.}
\end{figure}

We check the conditions $(\rm I)$ and $(\rm II)$ for the unitary operator $U(\tau)$.
Condition $(\rm I)$ is satisfied since the subspace spanned by
$\{U(\tau)|00\rangle,U(\tau)|01\rangle,U(\tau)|10\rangle,U(\tau)|11\rangle\}$ coincides with the subspace $\mathbf S=\mathrm{Span}\{|00\rangle,|01\rangle,|10\rangle,|11\rangle\}$. Furthermore, since
$\bar{H}$ commutes with its evolution operator $U(t)$, condition $(\rm II)$ reduces to $\langle k|\bar{H}|k'\rangle=0$, where $k,k'=\{00,01,10,11\}$.
 Thus, both conditions $(\rm I)$ and $(\rm II)$ are
satisfied and $U(\tau)$ is a holonomic two-qubit   swap gate in the subspace ${\mathbf S}$. 
{Fig.~\ref{p22} shows the temporal evolution of all ground states $\{|00\rangle,|01\rangle,|10\rangle,|11\rangle\}$ obtained from the full Hamiltonian of Eq.~(\ref{initial2}). The left panel of Fig.~\ref{p22} corresponds to the case for time-dependent control, $\Omega(t)=\Omega_{m}\mathrm{exp}[-(t-2T)^{2}/T^{2}]$ with $\Omega_{m}/(2\pi)=0.5$ MHz and
$T=\sqrt{\pi}/(\mathrm{Erf}[2]\Omega_{m})$, where the average value of $\Omega(t)$ is $\overline{\Omega(t)}/(2\pi) =0.22$~MHz.
Furthermore, for the case of time-independent  driving, the time-independent coupling coefficient $\Omega$, which governs the system’s time evolution, must be sufficiently small (e.g.  $\Delta\gg \Omega_{S}\gg \Omega$ and $\Delta=\sqrt{2}J$) to achieve
better addressing of resonant terms and efficient suppression of off-resonant interactions. Consequently, we set the time-independent  driving $\Omega/(2\pi) =0.1$ MHz in the right panel of Fig.~\ref{p22}. 
For both cases, the other parameters take the same values, which are $\Omega_{S} /(2\pi) =5$~MHz, $\Delta /(2\pi) =500 \sqrt{2}$ MHz, and $J/(2\pi)= 500$ MHz. Comparing the left and right panels of Fig.~\ref{p22}, it is shown that time-dependent control allows us to effectively mitigate the impact of off-resonant terms even at higher driving intensities (e.g. $\overline{\Omega(t)}/(2\pi) =0.22$~MHz and $\Omega/(2\pi) =0.1$ MHz), achieving this within a shorter time frame $\tau=2.267~\mu\rm s$ compared to time-independent driving, which requires the evolution time being $\tau'=5~\mu \rm s$.
}

\begin{figure}
\centering
\scalebox{0.5}{\includegraphics{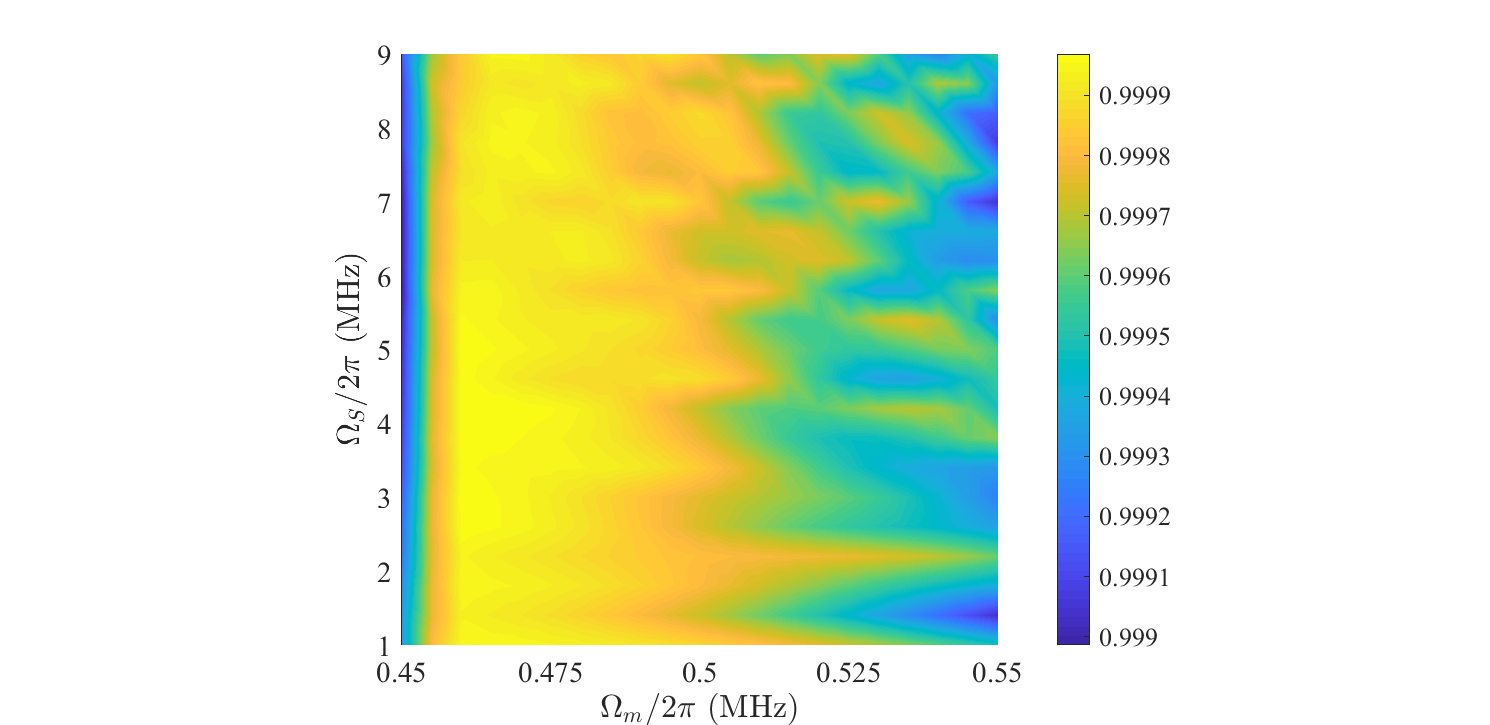}} \caption{\label{ppp3}
 Fidelity  versus the parameters $\Omega_{m}$ and $\Omega_{S}$  to the
target evolution without unwanted coupling by using a Gaussian soft coupling  $\Omega(t)=\Omega_{m}\mathrm{exp}[-(t-2T)^{2}/T^{2}]$.
 The other parameters are    
 $\Delta /(2\pi) =500 \sqrt{2}$ MHz, $J/(2\pi)= 500$ MHz, and $T=\sqrt{\pi}/(\mathrm{Erf}[2]\Omega_{m})$.}
\end{figure}

{The fidelity of the two-qubit swap gate in the
ideal case is $ F=\langle \psi_{{\rm ideal}} |\rho(t)| \psi_{{\rm ideal}}\rangle=99.98\%$ in the evolution period $\tau=2.267~\mu\rm s$ for the time-dependent control case, while for
the time-independent driving case, the fidelity is $F=99.97\%$ at the evolution period $\tau'=5~\mu\rm s$,
where the initial  state is $(|00\rangle + |01\rangle - |10\rangle + |11\rangle)/2$ and  the ideal final state is $(|00\rangle - |01\rangle +|10\rangle + |11\rangle)/2$.  
Hence, by employing the time-dependent
control, one can obtain a quantum gate with higher fidelity in a shorter time.
Fig.~\ref{ppp3} shows the fidelity of the swap gate versus the parameters $\Omega_{m}$ and $\Omega_{S}$. 
 An inspection of the plot reveals that the soft-coupling approach results in much higher fidelities in a
wide range of parameters, even for strong coupling regimes and a wide range of evolution times. 
Thus, we can say in the SRP mechanism based on the technique of soft quantum control that the two-qubit swap gate is implemented with high and stable
fidelity in a shorter time frame. }

\section{Performance of the holonomic swap gate}\label{sec3}

\subsection{Fluctuations of relevant parameters}\label{subsec2}
\begin{figure}
\centering
   \scalebox{0.48}{\includegraphics{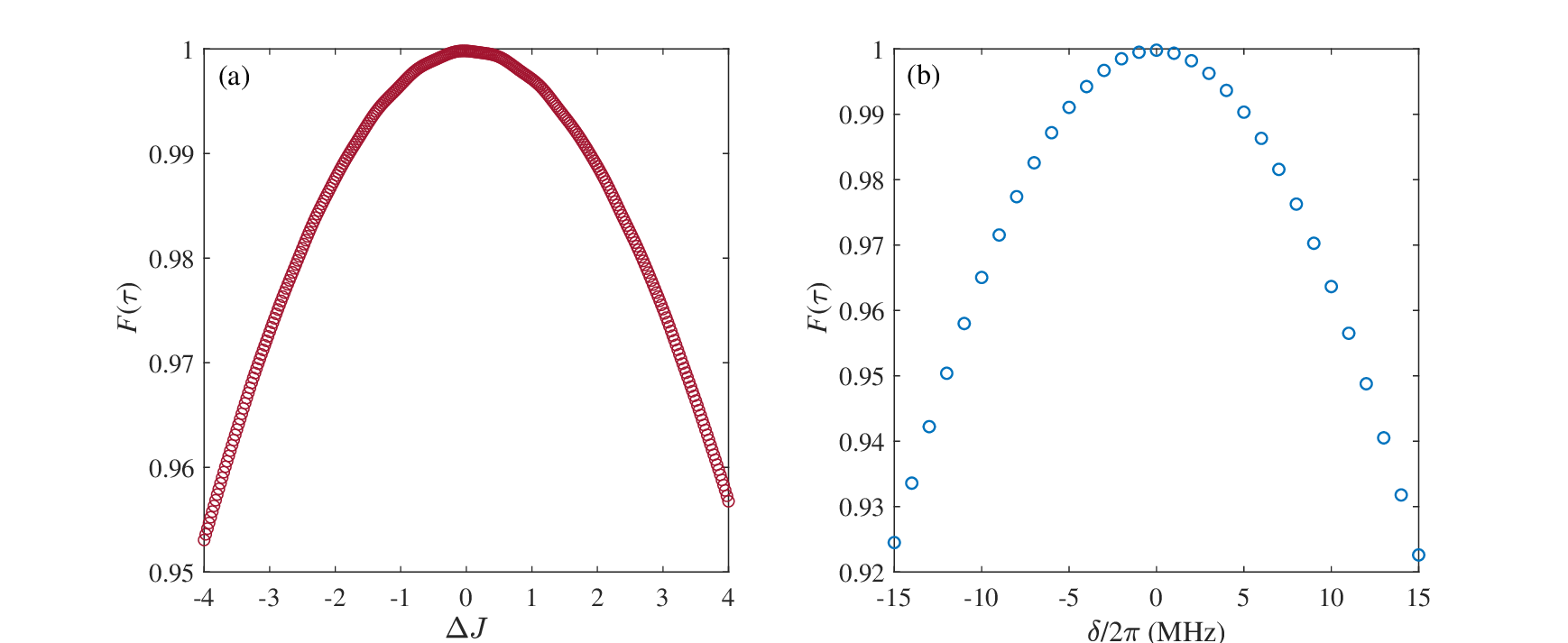}} \caption{\label{distance1}
 The effect of deviation $\Delta J$ (a) and the F\"{o}rster defect $\delta$ (b) on the fidelity of the swap gate based on the technique of soft quantum control. The  time-dependent parameter is $\Omega(t)=\Omega_{m}\mathrm{exp}[-(t-2T)^{2}/T^{2}]$ with $\Omega_{m}/(2\pi)=0.5$ MHz and $T=\sqrt{\pi}/(\mathrm{Erf}[2]\Omega_{m})$.  The other parameters
are  $\Omega_{S} /(2\pi) =5~$MHz, 
$\Delta/(2\pi)=500\sqrt{2}$~MHz, and $J/(2\pi)=(500+\Delta J)$~MHz.}
  \end{figure}

In the above process, we have assumed the detuning $\Delta =\sqrt{2}J$, which can be challenging to achieve precisely in experiments. To assess the impact of deviations from the desired dipole-dipole interaction on the SRP mechanism, 
we consider $J/(2\pi)=(500+\Delta J)$ MHz with the detuning parameter $\Delta/(2\pi) =500\sqrt{2}$ MHz. In Fig.~\ref{distance1}(a), the fidelity of the swap gate is plotted against the deviation $\Delta J$. Interestingly, the current SRP mechanism is shown to be insensitive to fluctuations in coupling strength $J$ between two atoms, since the fidelity of the gate consistently remains above $99\%$ in the continuous range of the coupling strength from $\Delta J=-1.78$ to $\Delta J=1.88$.
Furthermore, in Fig.~\ref{distance1}(b) we explore another practical scenario involving the presence of the F\"{o}rster defect.
In this case, the dipole-dipole coupling between the two Rydberg atoms in Eq.~(\ref{initial2})  is modified as $H_{dd}=J|dd\rangle(\langle pf_1|+\langle f_1p|)+\mathrm{H.c.}+\delta (|pf_1\rangle\langle pf_1|+|f_1p\rangle\langle f_1p|)$, where $\delta$ represents the F\"{o}rster defect. Surprisingly, the fidelity of the swap gate remains unaffected by the  F\"{o}rster defect, staying above $92\%$ throughout the continuous range of $\delta/(2\pi)=-15$ to $\delta/(2\pi)=15$ MHz.

  \subsection{Influence of spontaneous emission of Rydberg states}\label{subsec3}
When we consider the spontaneous emission of the Rydberg states, the Markovian master equation of the system in Lindblad form reads:
\begin{equation}\label{rho}
   \dot{\rho}=-i\big[H_{\mathrm{full}},\rho\big]+\sum_{k=1}^{2}\sum_{l=d,p,f_{1}}\{ \sum_{m=0}^{1} \gamma_{l}^{m} \mathcal{D}[|m\rangle_{k} \langle l|] +\sum_{n=a_{1}}^{a_{n}}\gamma_{l}^{n}\mathcal{D}[|n\rangle_{k} \langle l| ]\},
\end{equation}
where $\mathcal{D}[|m\rangle\langle l|]=[|m\rangle\langle l|\rho |l\rangle\langle m|-1/2(|l\rangle\langle l|\rho+\rho |l\rangle\langle l|)]$
and $\{|a_{1}\rangle,\cdots, |a_{n}\rangle\}$ denotes the subspace consists of the external leakage levels out of $\{|0\rangle, |1\rangle\}$ \cite{Levine,Madjarov}.
$\gamma_{l}^{m}$ is the branching ratio of the spontaneous decay rate from the state $|l\rangle$ to $|m\rangle$, which satisfies $\gamma_{l}=(\sum_{m=0}^{1} \gamma_{l}^{m}+\sum_{n=a_{1}}^{a_{n}}\gamma_{l}^{n})=1/\tau_{l}$. For the sake of simplifying the calculations, we let $\gamma_{l}^{m}=\gamma_{l}^{n}=\gamma_{l}/8$. The effective lifetimes of the Rydberg states $|p\rangle=|61 P_{1/2},m_{J}=1/2\rangle$, $|d\rangle=|59 D_{3/2},m_{J}=3/2\rangle$, and $|f_{1}\rangle=|57 F_{5/2},m_{J}=5/2\rangle$ of two $^{87}\mathrm{Rb}$ atoms
are $\tau_{p}=0.527$~ms, $\tau_{d}=0.215$~ms, and $\tau_{f_{1}}=0.127$~ms, respectively. In this case, the fidelity for the two-qubit swap gate is $ F=\langle \psi_{{\rm ideal}} |\rho(t)| \psi_{{\rm ideal}}\rangle=99.85\%$, where the initial  state is $(|00\rangle + |01\rangle - |10\rangle + |11\rangle)/2$ and  the ideal final state is $(|00\rangle - |01\rangle +|10\rangle + |11\rangle)/2$. Accordingly, the swap gate is shown to be robust against decoherence.

\section{ Holonomic three-qubit controlled-swap gate}\label{sec4}

\begin{figure}
\centering
\scalebox{0.5}{\includegraphics{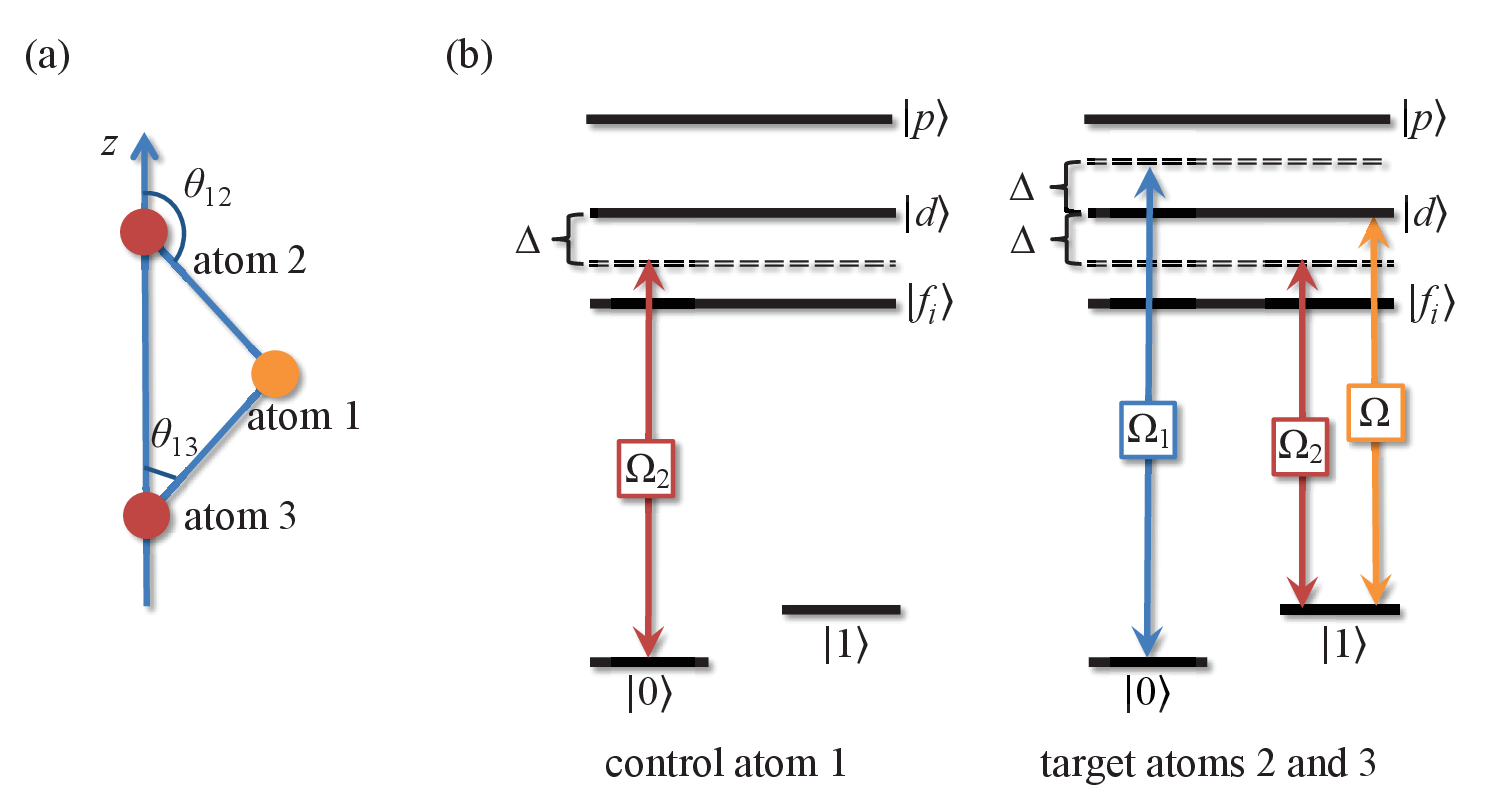}} \caption{\label{p1}
 Schematic view of the atomic-level configuration. (a) The arrangement of the control atom 1 and target atoms 2 and 3. (b) One laser field with the Rabi frequency $\Omega_{2}$ is applied to drive the transition $|0\rangle \leftrightarrow |d\rangle$ of the first atom (control atom)
with a red detuning  $\Delta$. Three types of laser fields are applied to drive the second and third atoms. One laser field with the Rabi frequency $\Omega_{1}$ is applied to drive the transition $|0\rangle \leftrightarrow |d\rangle$
with a blue detuning  $\Delta$. Meanwhile, the ground state $|1\rangle$ is dispersively coupled to the excited state $|d\rangle$  by one laser field
with Rabi frequency $\Omega_{2}$ (which has a red detuning $\Delta$) and another
resonant laser field with Rabi frequency $\Omega$, simultaneously.
 }
\end{figure}

In this section, we implement a holonomic controlled-swap gate by introducing an additional atom as the control atom, which has many known applications, such as preparation of the quantum state \cite{Ozaydin,Araujo,Murta}, quantum switches \cite{Chiribella,Baumeler,Castro}, and a variational quantum algorithm \cite{Preskill,Jones}.

The interaction between two neutral atoms $i$ and $j$ separated by a distance $R_{ij}$ can be expressed, to the leading order, through the dipole-dipole interaction
\cite{PhysRevA.62.052302,Barredo2015,Labuhn2016,Barredo2018,Ravon2023}:
\begin{equation}\label{Vddi}
V_{dd} = \frac{1}{4\pi\varepsilon_{0}}
\frac{\mathbf{d}_{i}\cdot\mathbf{d}_{j}-3(\mathbf{d}_{i}\cdot\mathbf{n}_{ij})(\mathbf{d}_{j}\cdot\mathbf{n}_{ij})}{R_{ij}^{3}},
\end{equation}
where $\mathbf{d}_{i}=(d_{x},d_{y},d_{z})$ is the electric dipole moment operator of atom $i$, and $\mathbf{n}_{ij} = \mathbf{R}_{ij}/R_{ij}$ is the unit vector connecting the two atoms from atoms $i$ to $j$. We denote the quantization axis with $z$, and the angle between $z$ and $\mathbf{n}_{ij}$ with $\theta_{ij}$.
In the spherical basis, it is convenient to use the spherical dipole operators: $d_{i,0}=d_{i,z}$ and
$d_{i,\pm}=\mp(d_{i,x}\pm id_{i,y})/\sqrt{2}$. The operator $d_{i,0}$ conserves the magnetic quantum number $m_{j}$, 
whereas the operators $d_{i,\pm}$ change $m_{j}$ by one ($\Delta m_{j}=\pm 1$).   In the spherical basis, the dipole-dipole interaction can be written as:
\begin{eqnarray}\label{Vdd}
V_{dd} &=& \frac{1}{2}\sum_{i\neq j}\frac{1}{4\pi\varepsilon_{0}R_{ij}^{3}}\big[\mathcal{A}_{1}(\theta_{ij})(d_{i,+}d_{j,-}+d_{i,-}d_{j,+}+2d_{i,z}d_{j,z})+\mathcal{A}_{2}(\theta_{ij})(d_{i,+}d_{j,z}\nonumber \\
&&-d_{i,-}d_{j,z}+d_{i,z}d_{j,+}-d_{i,z}d_{j,-})-\mathcal{A}_{3}(\theta_{ij})(d_{i,+}d_{j,+}+d_{i,-}d_{j,-})\big].
\end{eqnarray}
The operator $V_{dd}$ 
in Eq.~(\ref{Vdd}) contains three terms with angular prefactors
$\mathcal{A}_{1}(\theta_{ij})=(1-3\cos^{2}\theta_{ij})/2$, $\mathcal{A}_{2}(\theta_{ij})=3\sin\theta_{ij} \cos\theta_{ij}/\sqrt{2}$, and $\mathcal{A}_{3}(\theta_{ij})=3\sin^{2}\theta_{ij}/2$,
which couple pair states where the total magnetic quantum number $M=m_{1}+m_{2}$ changes by $\Delta M=0$, $\Delta M=\pm 1$, and $\Delta M=\pm 2$, respectively.
 Here, we consider the situation where no magnetic field is applied to the atoms. Coming back
to Eq.~(\ref{Vdd}), terms with the angular prefactor $\mathcal{A}_{1}(\theta_{ij})$ couple
$|dd\rangle$ and the symmetric state $|pf_{1}\rangle_{\rm s}=(|pf_{1}\rangle+|f_{1}p\rangle)/\sqrt{2}$, with $|f_{1}\rangle=|57 F_{5/2},m_{J}=5/2\rangle~(\Delta M=0)$. Terms with the angular prefactor $\mathcal{A}_{2}(\theta_{ij})$ couple $|dd\rangle$ and $|pf_{2}\rangle_{\rm s}=(|pf_{2}\rangle+|f_{2}p\rangle)/\sqrt{2}$, with $|f_{2}\rangle=|57 F_{5/2},m_{J}=3/2\rangle$ ($\Delta M=-1$). Finally, terms with the angular prefactor
$\mathcal{A}_{3}(\theta_{ij})$ couple $|dd\rangle$ and $|pf_{3}\rangle_{\rm s}=(|pf_{3}\rangle+|f_{3}p\rangle)/\sqrt{2}$, with $|f_{3}\rangle=|57 F_{5/2},m_{J}=1/2\rangle$ ($\Delta M=-2$). Thus, for the case of three atoms, each atom consists of
five long-lived Rydberg states $|d\rangle$, $|p\rangle$, $|f_{1}\rangle$, $|f_{2}\rangle$, and $|f_{3}\rangle$, and two ground states $|0\rangle$ and $|1\rangle$ shown in Fig.~\ref{p1}(b). Furthermore, the ground state $|0\rangle$ of the control atom is coupled to the excited state $|d\rangle$ by a laser field with Rabi frequency $\Omega_{2}$ (which has a red detuning $\Delta$). 
In the absence of a magnetic field, we thus expect three resonances between $|dd\rangle$ and the states $|pf_{1}\rangle_{\rm s}$, $|pf_{2}\rangle_{\rm s}$, and $|pf_{3}\rangle_{\rm s}$. The dipole-dipole interaction for the three-atom model has a concise form
\begin{eqnarray}\label{Vdd1}
V_{dd} &=& \sum_{i\neq j}\frac{1}{\sqrt{2}R_{ij}^3}\big[-\mathcal{A}_{1}(\theta_{ij})C_3|dd\rangle_{ij} \langle pf_{1}|_{\rm s}+\mathcal{A}_{2}(\theta_{ij})C'_3|dd\rangle_{ij} \langle pf_{2}|_{\rm s}\nonumber\\&&+\mathcal{A}_{3}(\theta_{ij})C''_3|dd\rangle_{ij} \langle pf_{3}|_{\rm s}\big]+\mathrm{H.c.},
\end{eqnarray}
where $C'_3/(2\pi)=1.61~{\rm GHz~\mu m}^3$ and $C''_3/(2\pi)=0.8~{\rm GHz~\mu m}^3$ \cite{SIBALIC2017319,Weber_2017}.
The arrangement of control atom 1 and target atoms 2 and 3 is shown in Fig.~\ref{p1}(a). We denote atoms 2 and 3 along the quantization $z$ axis, so the angle between $z$ and $\mathbf{n}_{23}$ is $\theta_{23}=0$. 
Meanwhile, the angle between $z$ and $\mathbf{n}_{13}$ is $\theta_{13}=\theta$, and the angle between $z$ and $\mathbf{n}_{12}$ is $\theta_{12}=\pi-\theta$.
We can fix the angle $\theta\approx55.80\degree$  to make the eigenvalues of the electric dipole-dipole interaction between any pair of three atoms equal, given that $4\cos^{3}\theta\sqrt{C^{2}_{3}(3\cos^{2}\theta-1)^{2}+18C'^{2}_{3}\sin^{2}\theta \cos^{2}\theta+9C''^{2}_{3}\sin^{4}\theta}/C_{3}=1$.

The Hamiltonian of the three-atom system in the interaction picture reads 
\begin{equation}\label{initial1223}
H'_{\mathrm{full}} =\Omega_{2}e^{-i\Delta t}|0\rangle_{1}\langle d|
+\sum_{k=2}^{3}\big[\Omega_{1}e^{i\Delta t}|0\rangle_{k}\langle d|+(\Omega+\Omega_{2}e^{-i\Delta t})|1\rangle_{k}\langle d|\big]+\mathrm{H.c.}+{V}_{dd}.
\end{equation}
Due to the strong dipole-dipole interaction strength between
the  Rydberg atoms, we make a rotation with respect to
{\begin{equation}
\label{transform2}
U_{\mathrm{rot}} =\exp\big[-\sqrt{2}iJt\sum_{k=1}^{6}(|E_{k+}\rangle\langle E_{k+}|-|E_{k-}\rangle\langle E_{k-}|)\big],
\end{equation}
 where 
\begin{eqnarray}\label{eigenstates}
|E_{1\pm}\rangle&=&\frac{1}{\sqrt{2}}\big(|0dd\rangle\pm|0pf_{1}\rangle_{\rm s}\big),\nonumber\\
|E_{2\pm}\rangle&=&\frac{1}{\sqrt{2}}\big[|d0d\rangle\pm\frac{2}{C_{3}}\big(\mathcal{B}_{1}|p0f_{1}\rangle_{\rm s}+\mathcal{B}_{2}|p0f_{2}\rangle_{\rm s}+\mathcal{B}_{3}|p0f_{3}\rangle_{\rm s}\big)\big],\nonumber\\
|E_{3\pm}\rangle&=&\frac{1}{\sqrt{2}}\big[|dd0\rangle\pm\frac{2}{C_{3}}\big(\mathcal{B}_{1}|pf_{1}0\rangle_{\rm s}-\mathcal{B}_{2}|pf_{2}0\rangle_{\rm s}+\mathcal{B}_{3}|pf_{3}0\rangle_{\rm s}\big)\big],\nonumber\\
|E_{4\pm}\rangle&=&\frac{1}{\sqrt{2}}\big(|1dd\rangle\pm|1pf_{1}\rangle_{\rm s}\big),\nonumber\\
|E_{5\pm}\rangle&=&\frac{1}{\sqrt{2}}\big[|d1d\rangle\pm\frac{2}{C_{3}}\big(\mathcal{B}_{1}|p1f_{1}\rangle_{\rm s}+\mathcal{B}_{2}|p1f_{2}\rangle_{\rm s}+\mathcal{B}_{3}|p1f_{3}\rangle_{\rm s}\big)\big],\nonumber\\
|E_{6\pm}\rangle&=&\frac{1}{\sqrt{2}}\big[|dd1\rangle\pm\frac{2}{C_{3}}\big(\mathcal{B}_{1}|pf_{1}1\rangle_{\rm s}-\mathcal{B}_{2}|pf_{2}1\rangle_{\rm s}+\mathcal{B}_{3}|pf_{3}1\rangle_{\rm s}\big)\big],\nonumber
\end{eqnarray}
are the eigenstates of the Rydberg dipole-dipole interaction with the eigenvalues being $ E_{k\pm}=\pm \sqrt{2}J$ ($k=1,2,\cdots,6$) respectively,  where $J=C_{3}/R^{3}_{23}$.} The symmetric
states are $|0pf_{1}\rangle_{\rm s}=(|0pf_{1}\rangle+|0f_{1}p\rangle)/\sqrt{2}$, $|p0f_{l}\rangle_{\rm s}=(|p0f_{l}\rangle+|f_{l}0p\rangle)/\sqrt{2}$, $|pf_{l}0\rangle_{\rm s}=(|pf_{l}0\rangle+|f_{l}p0\rangle)\sqrt{2}$, $|1pf_{1}\rangle_{\rm s}=(|1pf_{1}\rangle+|1f_{1}p\rangle)/\sqrt{2}$, $|p1f_{l}\rangle_{\rm s}=(|p1f_{l}\rangle+|f_{l}1p\rangle)/\sqrt{2}$, and $|pf_{l}1\rangle_{\rm s}=(|pf_{l}1\rangle+|f_{l}p1\rangle)\sqrt{2})/\sqrt{2}$ with $l=1,2,3$. 
The coefficients are $\mathcal{B}_{1}=2C_{3}\cos^{3}\theta(3\cos^{2}\theta-1)$, $\mathcal{B}_{2}=6\sqrt{2}C'_{3}\sin\theta \cos^{4}\theta$, and $\mathcal{B}_{3}=6C''_{3}\sin^{2}\theta \cos^{3}\theta$.
The  transformed Hamiltonian 
takes the following form
\begin{equation}
\label{initial13}
H'_{\mathrm{full}}=H'_{1}+H'_{2},
\end{equation}
\begin{eqnarray}
H'_{1}&=&\Omega_{S}|000\rangle(\langle 0d0|+\langle 00d|)e^{i\Delta t}+|001\rangle\big[\langle 00d|(\Omega+\Omega_{S}e^{-i\Delta t})+\langle 0d1|\Omega_{S}e^{i\Delta t}\big]\nonumber\\
&&+|010\rangle\big[\langle 0d0|(\Omega+\Omega_{S}e^{-i\Delta t})+\langle 01d|\Omega_{S}e^{i\Delta t}\big]+|011\rangle(\langle 01d|+\langle 0d1|)(\Omega+\Omega_{S}e^{-i\Delta t})\nonumber\\
&&+\sqrt{2}\Omega_{S}|100\rangle\langle T'_{0}|e^{i\Delta t}+\frac{1}{\sqrt{2}}|101\rangle\big[(\langle T'_{0}|-\langle S'_{0}|)(\Omega+\Omega_{S}e^{-i\Delta t})+\langle 1d1|\sqrt{2}\Omega_{S}e^{i\Delta t}\big]\nonumber\\
&&+\frac{1}{\sqrt{2}}|110\rangle\big[(\langle T'_{0}|+\langle S'_{0}|)(\Omega+\Omega_{S}e^{-i\Delta t})+\langle 11d|\sqrt{2}\Omega_{S}e^{i\Delta t}\big]\nonumber\\
&&+\sqrt{2}(\Omega+\Omega_{S}e^{-i\Delta t})|111\rangle\langle T'_{1}|+\mathrm{H.c.} ,\nonumber
\end{eqnarray}
\begin{eqnarray}
H'_{2}&=&\frac{1}{\sqrt{2}}\Omega_{S}|00d\rangle\big[\langle E_{1+}|e^{i(\Delta-\sqrt{2}J)t}+\langle E_{1-}|e^{i(\Delta+\sqrt{2}J)t}+\langle E_{2+}|e^{-i(\Delta+\sqrt{2}J) t}\nonumber\\
&&+\langle E_{2-}|e^{-i(\Delta-\sqrt{2}J) t}\big]+\frac{1}{\sqrt{2}}\Omega_{S}|0d0\rangle\big[\langle E_{1+}|e^{i(\Delta-\sqrt{2}J)t}+\langle E_{1-}|e^{i(\Delta+\sqrt{2}J)t}\nonumber\\
&&+\langle E_{3+}|e^{-i(\Delta+\sqrt{2}J) t}+\langle E_{3-}|e^{-i(\Delta-\sqrt{2}J) t}\big]+\frac{1}{{\sqrt{2}}}|01d\rangle\big[\langle E_{1+}|(\Omega e^{-i\sqrt{2}J t}
\nonumber\\
&&+\Omega_{S}e^{-i(\Delta+\sqrt{2}J) t})+\langle E_{1-}|(\Omega e^{i\sqrt{2}J t}+\Omega_{S}e^{-i(\Delta-\sqrt{2}J) t})+\langle E_{5+}|\Omega_{S}e^{-i(\Delta+\sqrt{2}J) t}\nonumber\\
&&+\langle E_{5-}|\Omega_{S}e^{-i(\Delta-\sqrt{2}J) t}\big]+\frac{1}{\sqrt{2}}|0d1\rangle\big[\langle E_{1+}|(\Omega e^{-i\sqrt{2}J t}+\Omega_{S}e^{-i(\Delta+\sqrt{2}J) t})\nonumber\\
&&+\langle E_{1-}|(\Omega e^{i\sqrt{2}J t}+\Omega_{S}e^{-i(\Delta-\sqrt{2}J) t})+\langle E_{6+}|\Omega_{S}e^{-i(\Delta+\sqrt{2}J) t}\nonumber\\
&&+\langle E_{6-}|\Omega_{S}e^{-i(\Delta-\sqrt{2}J) t}\big]+\Omega_{S}|T'_{0}\rangle\big[\langle E_{4+}|e^{i(\Delta-\sqrt{2}J)t}+\langle E_{4-}|e^{i(\Delta+\sqrt{2}J)t}\big]\nonumber\\
&&+|T'_{1}\rangle\big[\langle E_{4+}|(\Omega e^{-i\sqrt{2}J t}+\Omega_{S}e^{-i(\Delta+\sqrt{2}J) t})+\langle E_{4 -}|(\Omega e^{i\sqrt{2}J t}+\Omega_{S}e^{-i(\Delta-\sqrt{2}J) t})\big]\nonumber\\
&&+\mathrm{H.c.},\nonumber
\end{eqnarray} 
where $\Omega_{1}=\Omega_{2}=\Omega_{S}$ for simplicity.  $|T'_{0}(S'_{0})\rangle=(|1d0\rangle\pm|10d\rangle)/\sqrt{2}$ and $|T'_{1}\rangle=(|1d1\rangle+|11d\rangle)/\sqrt{2}$. As is the case for two-atom,  we also consider  the large detuning case  (e.g. $\Delta\gg \{\Omega_{S},\Omega\}$ and  $\Delta=\sqrt{2}J$),  the Hamiltonian  can be reduced as 

\begin{eqnarray}\label{initial14}
H' &=&\Omega\big[|001\rangle\langle 00d|+|010\rangle\langle 0d0|+|011\rangle(\langle 01d|
+\langle 0d1|)\big]+\frac{1}{\sqrt{2}}\Omega(|110\rangle-|101\rangle)\langle S'_{0}|\nonumber\\
&&+\frac{1}{\sqrt{2}}\Omega(|110\rangle+|101\rangle)\langle T'_{0}|+\sqrt{2}\Omega|111\rangle\langle T'_{1}|
+\frac{1}{\sqrt{2}}\Omega_{S}|00d\rangle(\langle E_{1+}|+\langle E_{2-}|)\nonumber\\
&&
+\frac{1}{\sqrt{2}}\Omega_{S}|0d0\rangle(\langle E_{1+}|+\langle E_{3-}|)+\frac{1}{\sqrt{2}}\Omega_{S}|01d\rangle(\langle E_{1-}|+\langle E_{5-}|)\nonumber\\
&&
+\frac{1}{\sqrt{2}}\Omega_{S}|0d1\rangle(\langle E_{1-}|+\langle E_{6-}|)
+\Omega_{S}|T'_{0}\rangle\langle E_{4+}|+\Omega_{S}|T'_{1}\rangle\langle E_{4-}|
+\mathrm{H.c.}.
\end{eqnarray}

Similarly, we employ the  soft quantum control  $\Omega(t)=\Omega_{m}\mathrm{exp}[-(t-2T)^{2}/T^{2}]$ to the three-atom system. The Hamiltonian $H'$ in Eq.~(\ref{initial14}) can be divided into two parts
\begin{equation}\label{initial22222}
H' =H'_{\mathrm{S}}+H'_{\mathrm{int}},
\end{equation}
where
\begin{eqnarray}\label{initial2222}
H'_{\mathrm{S}} &=&\frac{1}{\sqrt{2}}\Omega_{S}|00d\rangle(\langle E_{1+}|+\langle E_{2-}|)
+\frac{1}{\sqrt{2}}\Omega_{S}|0d0\rangle(\langle E_{1+}|+\langle E_{3-}|)\nonumber\\
&&+\frac{1}{\sqrt{2}}\Omega_{S}|01d\rangle(\langle E_{1-}|+\langle E_{5-}|)
+\frac{1}{\sqrt{2}}\Omega_{S}|0d1\rangle(\langle E_{1-}|+\langle E_{6-}|)\nonumber\\
&&+\Omega_{S}|T'_{0}\rangle\langle E_{4+}|+\Omega_{S}|T'_{1}\rangle\langle E_{4-}|
+\mathrm{H.c.},\nonumber\\
H'_{\mathrm{int}} &=&
\Omega\big[|001\rangle\langle 00d|+|010\rangle\langle 0d0|+|011\rangle(\langle 01d|
+\langle 0d1|)\big]+\frac{1}{\sqrt{2}}\Omega(|110\rangle-|101\rangle)\langle S'_{0}|\nonumber\\
&&+\frac{1}{\sqrt{2}}\Omega(t)(|110\rangle+|101\rangle)\langle T'_{0}|+\sqrt{2}\Omega(t)|111\rangle\langle T'_{1}|+\mathrm{H.c.}.\nonumber
\end{eqnarray}
Using the above same method from Eq.~(\ref{initial222}) to Eq.~(\ref{average1}), the corresponding average Hamiltonian of Eq.~(\ref{initial14}) is 
\begin{eqnarray}
  \bar{H}'&=&\frac{g}{\sqrt{2}}(|110\rangle-|101\rangle)\langle S'_{0}|+\frac{g}{\sqrt{2}} |S'_{0}\rangle(\langle110|-\langle101|)\nonumber\\
  &&+\frac{1}{4T}\int_{0}^{4T}\sqrt{\Omega(t)^{2}+\Omega_{S}^{2}} dt (|\phi_{1+}\rangle\langle \phi_{1+}|-|\phi_{1-}\rangle\langle \phi_{1-}|)
  \nonumber\\
&&+\frac{1}{4T}\int_{0}^{4T}\sqrt{2\Omega(t)^{2}+\Omega_{S}^{2}} dt ( |\phi_{2+}\rangle\langle \phi_{2+}|- |\phi_{2-}\rangle\langle\phi_{2-})|\nonumber\\
  &&+\frac{1}{4T}\int_{0}^{4T}\sqrt{\frac{2\Omega(t)^{2}+\Omega_{S}^{2}}{2}} dt (|\phi_{3+}\rangle\langle \phi_{3+}|- |\phi_{3-}\rangle\langle \phi_{3-}|)
 \nonumber\\
&& +\frac{1}{4T}\int_{0}^{4T}\sqrt{\frac{2\Omega(t)^{2}+3\Omega_{S}^{2}}{2}} dt (|\phi_{4+}\rangle\langle \phi_{4+}|-|\phi_{4-}\rangle\langle\phi_{4-}|)\nonumber\\
  &&+\frac{1}{4T}\int_{0}^{4T}\frac{\Omega_{S}}{\sqrt{2}} dt (|\phi_{5+}\rangle\langle\phi_{5+}|- |\phi_{5-}\rangle\langle\phi_{5-}|)
  \nonumber\\
&& +\frac{1}{4T}\int_{0}^{4T} \sqrt{\frac{4\Omega(t)^{2}+3\Omega_{S}^{2}}{2}} dt (|\phi_{6+}\rangle\langle\phi_{6+}|-|\phi_{6-}\rangle\langle\phi_{6-}|),
  \end{eqnarray}
where $g=\sqrt{\pi}\Omega_{m}\mathrm{Erf}[2]/4$. 
$|\phi_{1\pm}\rangle=(|T'_{0}\rangle\pm|E_{4+}\rangle)/\sqrt{2}$, $|\phi_{2\pm}\rangle=(|T'_{1}\rangle\pm|E_{4-}\rangle)/\sqrt{2}$,
$|\phi_{3\pm}\rangle=[|E_{3-}\rangle-|E_{2-}\rangle\pm(|0d0\rangle-|00d\rangle)]/2$,
$|\phi_{4\pm}\rangle=[|E_{3-}\rangle+|E_{2-}\rangle+2|E_{1+}\rangle\pm \sqrt{3}(|0d0\rangle+|00d\rangle)]/2\sqrt{3}$,
$|\phi_{5\pm}\rangle=[|E_{6-}\rangle-|E_{5-}\rangle\pm(|0d1\rangle-|01d\rangle)]/2$, and
$|\phi_{6\pm}\rangle=[|E_{6-}\rangle+|E_{5-}\rangle+2|E_{1-}\rangle\pm \sqrt{3}(|0d1\rangle+|01d\rangle)]/2\sqrt{3}$
are the  eigenstates of $H'_{\mathrm{S}}$ governed by $\Omega_{S}$.

\begin{figure}
\centering
\includegraphics[width=1\linewidth]{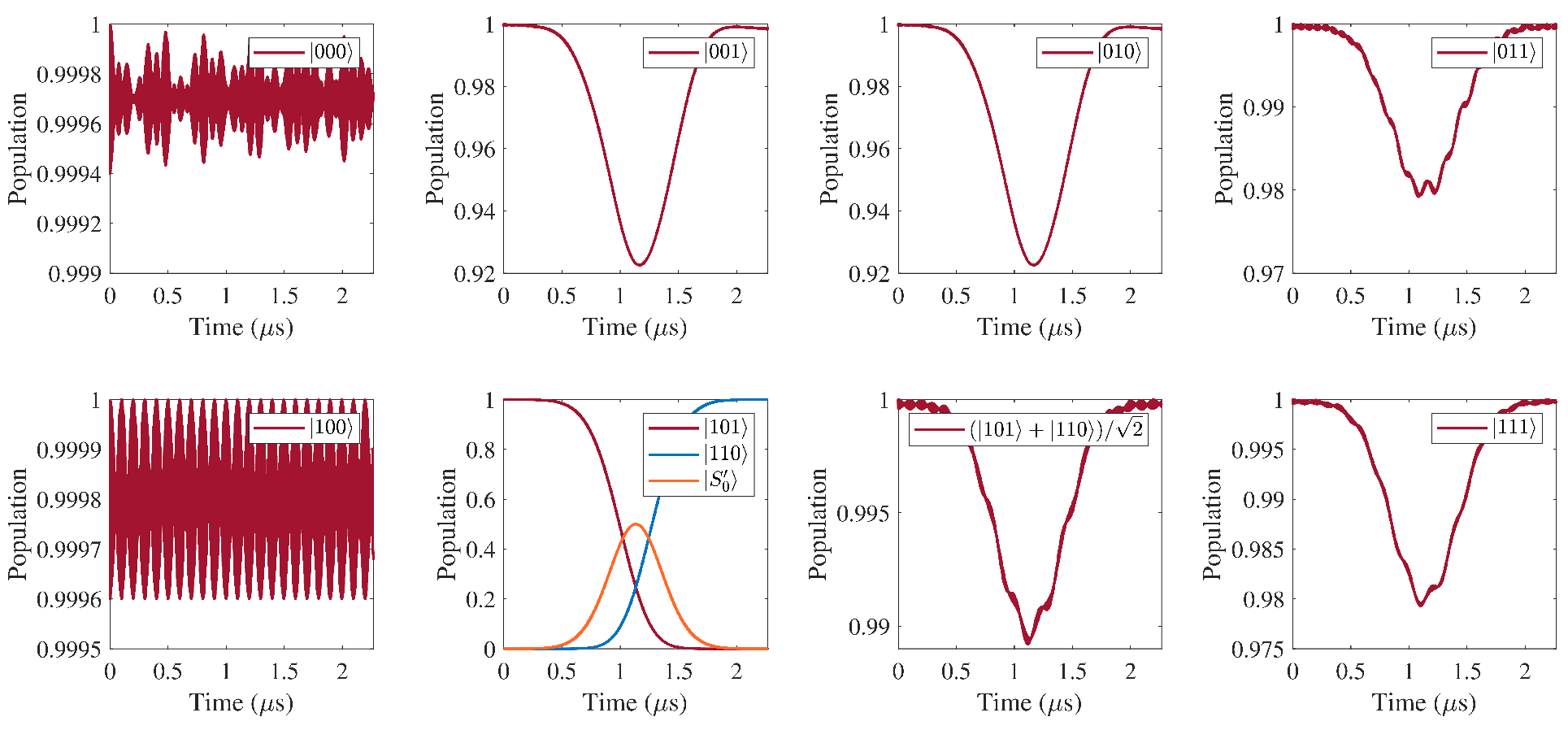}\caption{\label{p222}
 The temporal evolution of populations for different ground states governed by the full Hamiltonian in Eq.~(\ref{initial1223}).
The  time-dependent parameter is $\Omega(t)=\Omega_{m}\mathrm{exp}[-(t-2T)^{2}/T^{2}]$ with $\Omega_{m}/(2\pi)=0.5$ MHz and
$T=\sqrt{\pi}/(\mathrm{Erf}[2]\Omega_{m})$.  The other parameters are $\Omega_{S} /(2\pi) =5$ MHz, 
 $\Delta /(2\pi) =500\sqrt{2}$ MHz, and $J/(2\pi)= 500$ MHz.}
\end{figure}

The propagator $U'=e^{-4i\bar{H}' T}$ with the
evolution period $\tau=4T$ can generate a high-fidelity three-qubit controlled-swap gate. The  evolution operator $U'$ in the basis $\{|S'_{0}\rangle,|000\rangle,|001\rangle,|010\rangle,|011\rangle,|100\rangle,|101\rangle,|110\rangle,|111\rangle\}$ takes the following form 
\begin{eqnarray}
U'(\tau)=\left(\begin{array}{c c c c c c c c c}
\cos\lambda_{\tau}& 0 & 0& 0 & 0 & 0 &\frac{\sqrt{2}i}{2}\sin\lambda_{\tau} &-\frac{\sqrt{2}i}{2}\sin\lambda_{\tau} &0 \\
0 &1 & 0 & 0& 0 &0 &0 &0 &0\\
0& 0 &1 & 0 & 0 & 0 & 0 & 0 & 0\\
0 & 0 & 0 &1 & 0 & 0 & 0 &0 &0\\
0 & 0 & 0 & 0 & 1 & 0 & 0 &0 & 0 \\
0 & 0& 0& 0 & 0& 1& 0& 0& 0 \\
\frac{\sqrt{2}i}{2}\sin\lambda_{\tau} & 0& 0 &0 & 0 &0 & \frac{1}{2}(1+\cos\lambda_{\tau}) &\frac{1}{2}(1-\cos\lambda_{\tau}) & 0 \\
-\frac{\sqrt{2}i}{2}\sin\lambda_{\tau} & 0 & 0 & 0& 0 &0 & \frac{1}{2}(1-\cos\lambda_{\tau}) & \frac{1}{2}(1+\cos\lambda_{\tau}) &0 \\
 0 & 0& 0& 0& 0 & 0 &0 &0 & 1 \\
\end{array}
\right),\nonumber
\end{eqnarray}
where $\lambda_{\tau}=4Tg$.
Setting 
$
\lambda_{\tau}=4Tg=\pi
$,
 the final effective  evolution operator $U'(\tau)$ reads
 \begin{eqnarray}
U'(\tau)=\left(\begin{array}{c c c c c c c c c}
-1& 0& 0& 0 &0 & 0 &0 &0 & 0 \\
0& 1& 0& 0 &0 & 0 &0 &0 & 0 \\
0& 0& 1& 0 &0 & 0 &0 &0 & 0 \\
0& 0& 0& 1 &0 & 0 &0 &0 & 0 \\
0& 0& 0& 0 &1 & 0 & 0 &0 & 0 \\
0& 0& 0& 0 &0& 1& 0& 0& 0 \\
0& 0& 0& 0 &0 &0 & 0 &1 & 0 \\
0& 0& 0& 0 &0 &0 & 1 & 0 &0 \\
0& 0& 0& 0 &0 & 0 &0 &0 & 1 \\
\end{array}
\right),
\end{eqnarray}
which is a  three-qubit controlled-swap gate, in which the two target qubits swap their information $|01\rangle_{23}\Longleftrightarrow |10\rangle_{23}$ if and only if the control qubit is in $|1\rangle_{1}$, on the computational
subspace  $\mathbf S'=\mathrm{Span}\{|000\rangle,|001\rangle,|010\rangle,|011\rangle,|100\rangle,|101\rangle,|110\rangle,|111\rangle\}$ as follows
 \begin{eqnarray}
U'_{\mathrm{cswap}}=\left(\begin{array}{c c c c c c c c}
1& 0& 0 &0 & 0 &0 &0 & 0 \\
0& 1& 0 &0 & 0 &0 &0 & 0 \\
0& 0& 1 &0 & 0 &0 &0 & 0 \\
0& 0& 0 &1 & 0 & 0 &0 & 0 \\
0& 0& 0 &0& 1& 0& 0& 0 \\
0& 0& 0 &0 &0 & 0 &1 & 0 \\
0& 0& 0 &0 &0 & 1 & 0 &0 \\
0& 0& 0 &0 & 0 &0 &0 & 1 \\
\end{array}
\right).
\end{eqnarray}
 Using the same method as the holonomic proof for two-qubit swap gate, one can also confirm that both conditions $(\rm I)$ and $(\rm II)$ in Eq.~(\ref{condition}) are
satisfied. 
Therefore, $U'(\tau)$ is a holonomic three-qubit controlled-swap gate in subspace $\mathbf S'$.
Through the temporal evolution of all ground states obtained from the full Hamiltonian Eq.~(\ref{initial1223}) depicted in Fig.~\ref{p222}, it is shown that the controlled-swap gate $U'(\tau)$ within the subspace $\mathbf S'$ is pure holonomic. Additionally, the figure illustrates that  Gaussian time-dependent soft-control  enables efficient rotating-wave approximation across a wide parameter range.

The fidelity for the three-qubit controlled-swap gate in the ideal case is $ F=\langle \psi'_{{\rm ideal}}|\rho(t)| \psi'_{{\rm ideal}}\rangle=99.93\%$ after the same evolution period $\tau=2.267~\mu \rm s$ as for the two-qubit case. Here, the initial  state is  $(|000\rangle+|001\rangle+|010\rangle+|011\rangle+|100\rangle  + |111\rangle+\sqrt{3}|101\rangle-\sqrt{3}|110\rangle)/(2\sqrt{3})$ and the ideal final state is $(|000\rangle+|001\rangle+|010\rangle+|011\rangle+|100\rangle  + |111\rangle-\sqrt{3}|101\rangle+\sqrt{3}|110\rangle)/(2\sqrt{3})$ .

\begin{figure}
\centering
\scalebox{0.5}{\includegraphics{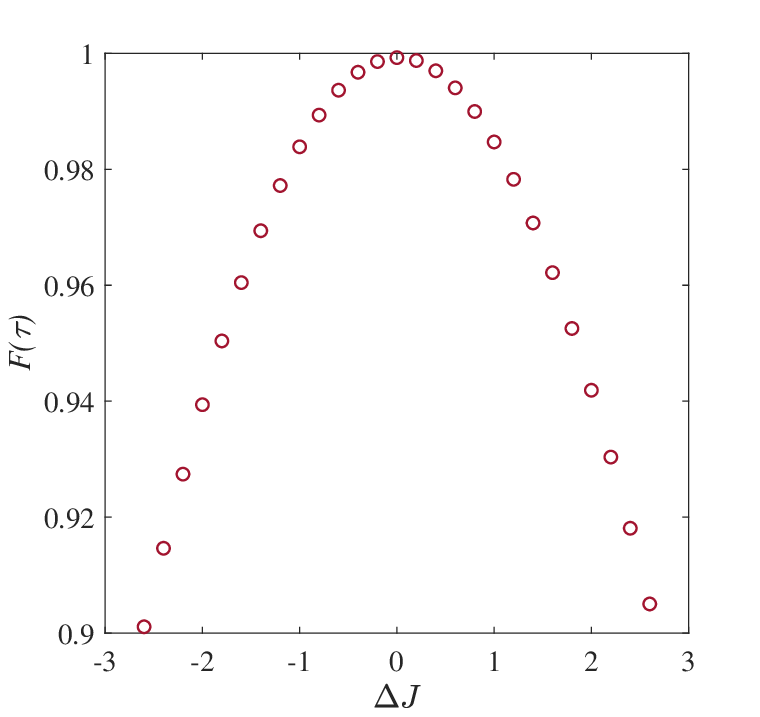}} \caption{\label{pjj}
The effect of deviation $\Delta J$ on the fidelity of the controlled-swap gate in the presence of F\"{o}rster defect $\delta/(2\pi) =8.5$ MHz.
The  time-dependent parameter is $\Omega(t)=\Omega_{m}\mathrm{exp}[-(t-2T)^{2}/T^{2}]$ with $\Omega_{m}/(2\pi)=0.5$ MHz and $T=\sqrt{\pi}/(\mathrm{Erf}[2]\Omega_{m})$.  The other parameters
are  $\Omega_{S} /(2\pi) =5~$MHz,  $\Delta_{1}/(2\pi) =711.37$~MHz , $\Delta_{2}/(2\pi) =702.87$~MHz, and $J/(2\pi)=(500+\Delta J)$~MHz.}
\end{figure}

{
We then discuss the practical situation in which F\"{o}rster defects exist. The pair states $|pf_l\rangle$ and $|f_lp\rangle$ are  degenerate with $l=1,2,3$, and  the 
F\"{o}rster defect between $|dd\rangle$ and $|pf_l\rangle$ ($|f_lp\rangle$) 
 is $\delta/(2\pi)=8.5$~$\mathrm{MHz}$ in the absence of an electric field~\cite{SR}. In this case, the dipole-dipole coupling between the three-atom in Eq.~(\ref{initial1223}) is modified as $H'_{dd}=V_{ dd}+\delta[|pf_{1}\rangle_{23}\langle pf_{1}|+|f_{1}p\rangle_{23}\langle f_{1}p|+\sum_{l=1,2,3}(|pf_{l}\rangle_{12}\langle pf_{l}|+|f_{l}p\rangle_{12}\langle f_{l}p|+|pf_{l}\rangle_{13}\langle pf_{l}|+|f_{l}p\rangle_{13}\langle f_{l}p|)]$. By making slight adjustments to the laser fields in Fig.~\ref{p1}, for the  control atom 1 we set the  laser field with the Rabi frequency $\Omega_{2}$ 
possess
 a red detuning  $\Delta_{2}=(\sqrt{8J^{2}+\delta^{2}}-\delta)/2$. Meanwhile, for target atoms 2 and 3 we set one laser field with the Rabi frequency $\Omega_{1}$ 
has a blue detuning  $\Delta_{1}=(\sqrt{8J^{2}+\delta^{2}}+\delta)/2$, and 
the other laser field with the Rabi frequency $\Omega_{2}$  
has a red detuning  $\Delta_{2}=(\sqrt{8J^{2}+\delta^{2}}-\delta)/2$. Through numerical simulations, it is found that the fidelity of the quantum gate is still  maintained at 99.93\%.
In the following, to assess the effect of deviations from the expected dipole-dipole interactions in the same F\"{o}rster defect, 
we consider $J/(2\pi)=(500+\Delta J)$ MHz with the detuning parameters $\Delta_{1}/(2\pi)= (\sqrt{8\cdot500^{2}+8.5^{2}}+8.5)/2=711.37$ MHz and $\Delta_{2}/(2\pi) = (\sqrt{8\cdot500^{2}+8.5^{2}}-8.5)/2=702.87$ MHz. In Fig.~\ref{pjj}, the fidelity of the controlled-swap gate is plotted against the deviation $\Delta J$, which  consistently remains  above $90\%$ in the continuous range of the coupling strength from $\Delta J=-2.6$ to $\Delta J=2.6$. Therefore, the current SRP mechanism is also insensitive to fluctuations in the coupling strength $J$  for three-qubit case.}

{We also consider the spontaneous emission of the Rydberg states  in the same F\"{o}rster defect. Based on the Markovian
master equation of the system in Lindblad form Eq.~(\ref{rho}), however, one needs to consider two extra Rydberg states $|f_{2}\rangle$ and $|f_{3}\rangle$ when the ploar angle $\theta_{ij}\neq 0$.
The fidelity for the three-qubit controlled-swap gate  is $ F=\langle \psi'_{{\rm ideal}} |\rho(t)| \psi'_{{\rm ideal}}\rangle=99.78\%$.
Consequently,  the three-qubit controlled-swap gate is implemented with a robust fidelity in a shorter time frame.}

\section{Conclusion}\label{sec6}

In conclusion, we have demonstrated a rapid implementation of holonomic swap and controlled-swap gates for neutral atoms using SRP. By incorporating  time-dependent control which enables highly selective coupling between different on-resonance constituents of composite quantum systems, within the SRP mechanism, we achieved an average Hamiltonian that achieves a higher and more stable population in a shorter time, leading to an efficient rotating-wave approximation across a broad parameter range. Our approach accelerates the synthesis of a robust two-qubit swap gate, which is robust against variations in the dipole-dipole interaction, F\"{o}rster defect fluctuation, and spontaneous emission of Rydberg states. Furthermore, our mechanism readily extends to the direct implementation of a holonomic three-qubit controlled-swap gate by introducing a control atom and selecting an appropriate angle $\theta$ between the interatomic axis and the quantization axis $z$ and an appropriate driving field. Combining the robustness against control imprecisions and high-speed evolution of nonadiabatic HQC, we hope our work may provide an
alternative approach toward fault-tolerant quantum computation.


\subsection*{Funding}

CFS is supported by the Plan for Scientific and Technological Development of Jilin Province (20240101315JC) and the scientific research project of the Education Department of Jilin Province (Grant No. JJKH20231293KJ).
XQS is supported by the National Natural Science Foundation (Grant No. 12174048).
GCW is supported by the scientific research project of the
Education Department of Jilin Province (Grants No. JJKH20241408KJ).
JBY acknowledges support from
the National Research Foundation Singapore (NRF2021-
QEP2-02-P01), A*STAR Career Development
Award (C210112010), and A*STAR (C230917003, C230917007).
\section*{Declarations}

\subsection*{Ethics approval and consent to participate}
Not applicable.

\subsection*{Consent for publication}
Not applicable.

\subsection*{Competing interests}
The authors declare no competing interests.

\subsection*{Author contributions}
XQS conceived the idea and designed the scheme. CFS and XYC derived the theoretical framework.  WLM and GCW implemented numerical simulations. CFS and XYC were the main contributors in the writing of the manuscript. XQS and JBY contributed to the comment and revision of the manuscript. All authors read and approved the final manuscript.

\subsection*{Author details}
${^1}$Center for Quantum Sciences and School of Physics, Northeast
Normal University, Changchun, 130024, China.
${^2}$Key Laboratory for UV Light-Emitting Materials and Technology of
Ministry of Education, Northeast Normal University, Changchun,
130024, China.
${^3}$Department of Physics, Beijing Normal University, Beijing 100875, China
${^4}$Institute of High-Performance Computing, A*STAR (Agency for
Science, Technology and Research), 1 Fusionopolis Way, Connexis,
138632, Singapore.

{\subsection*{Appendix A: The terms oscillating with high frequencies}

Under the conditons
$\Delta\gg \{\Omega_{S},\Omega(t)\}$  and  $\Delta=\sqrt{2}J$,
the Hamiltonian in Eq. (\ref{initial22}) can be written in the follwing form
\begin{eqnarray}\label{initial3}
H_{\rm full}&=&H_{\rm res}+H_{\rm dis},\\
H_{\rm res}&=&\frac{1}{\sqrt{2}}\Omega(t)(|10\rangle-|01\rangle)\langle S_{0}|+\frac{1}{\sqrt{2}}\Omega(t)(|10\rangle+|01\rangle)\langle T_{0}|+\sqrt{2}\Omega(t)|11\rangle\langle T_{1}|\nonumber\\
&&
+\Omega_{S}|T_{0}\rangle\langle E_{+}|+\Omega_{S}|T_{1}\rangle\langle E_{-}|+\mathrm{H.c.},\nonumber\\
H_{\rm dis}&=&\sqrt{2}\Omega_{S}|00\rangle\langle T_{0}|e^{i\Delta t}
+\frac{1}{\sqrt{2}}\Omega_{S}e^{-i\Delta t}(|10\rangle-|01\rangle)\langle S_{0}|\nonumber\\
&&+\frac{1}{\sqrt{2}}\Omega_{S}e^{-i\Delta t}(|10\rangle+|01\rangle)\langle T_{0}|+|01\rangle\langle d1|\Omega_{S}e^{i\Delta t}+|10\rangle\langle 1d|\Omega_{S}e^{i\Delta t}\nonumber\\
&&+\sqrt{2}\Omega_{S}e^{-i\Delta t}|11\rangle\langle T_{1}|+\Omega_{S}|T_{0}\rangle\langle E_{-}|e^{i2\Delta t}+|T_{1}\rangle\big[\langle E_{+}|(\Omega e^{-i\Delta t}
+\Omega_{S}e^{-i2\Delta t})\nonumber\\
&&+\langle E_{-}|\Omega e^{i\Delta t}\big]
+\mathrm{H.c.}.\nonumber
\end{eqnarray}
The terms in $H_{\rm dis}$ are the dispersive interaction and their actions are equal to the Stark shifts of atomic levels, and these Stark shifts can be canceled by introducing the other ancillary levels and laser fields to induce the opposite Stark shifts. Thus, the Hamiltonian in Eq. (2) can be evaluated
explicitly
\begin{eqnarray}\label{initial4}
H=H_{\rm res}&=&\frac{1}{\sqrt{2}}\Omega(t)(|10\rangle-|01\rangle)\langle S_{0}|+\frac{1}{\sqrt{2}}\Omega(t)(|10\rangle+|01\rangle)\langle T_{0}|+\sqrt{2}\Omega(t)|11\rangle\langle T_{1}|\nonumber\\
&&
+\Omega_{S}|T_{0}\rangle\langle E_{+}|+\Omega_{S}|T_{1}\rangle\langle E_{-}|+\mathrm{H.c.}.
\end{eqnarray}

}

{\subsection*{Appendix B: The average Hamiltonian}

The Hamiltonian $H_{\mathrm{S}}$ in Eq. ~(\ref{initial222}) has the eigenstates 
\begin{eqnarray}
|\psi^{D}_{1}\rangle&=&\frac{1}{\sqrt{2}}(|T_{0}\rangle+|E_{+}\rangle),\nonumber\\
|\psi^{D}_{2}\rangle&=&\frac{1}{\sqrt{2}}(|T_{0}\rangle-|E_{+}\rangle),\nonumber\\
|\psi^{D}_{3}\rangle&=&\frac{1}{\sqrt{2}}(|T_{1}\rangle+|E_{-}\rangle),\nonumber\\
|\psi^{D}_{4}\rangle&=&\frac{1}{\sqrt{2}}(|T_{1}\rangle-|E_{-}\rangle),\nonumber\\  |\psi^{D}_{5}\rangle&=&\frac{1}{\sqrt{2}}[|S_{0}\rangle+\frac{1}{\sqrt{2}}(|10\rangle-|01\rangle)],\nonumber\\  
|\psi^{D}_{6}\rangle&=&\frac{1}{\sqrt{2}}[|S_{0}\rangle-\frac{1}{\sqrt{2}}(|10\rangle-|01\rangle)],\nonumber\\    |\psi^{D}_{7}\rangle&=&\frac{1}{\sqrt{2}}\big(|10\rangle+|01\rangle\big),\nonumber\\     |\psi^{D}_{8}\rangle&=&|11\rangle.\nonumber
\end{eqnarray}
The corresponding eigenvalues are $\omega_{1}=\Omega_{S},\omega_{2}=-\Omega_{S},\omega_{3}=\Omega_{S},\omega_{4}=-\Omega_{S},\omega_{5}=\omega_{6}=\omega_{7}=\omega_{8}=0$ and the corresponding 
projection operators are
$\mathbb{P}(\omega_{j})=|\psi^{D}_{j}\rangle\langle\psi^{D}_{j}|$ ($j=1,2,...,8$) respectively.

The whole Hamiltonian   in Eq.~(\ref{initial222}) under the projection operators  $\mathbb{P}(\omega_{j})$ is written as $H^{1}=H_{{\rm S}}^{1}+H_{\mathrm{int}}^{1}$, and its form in the basis \{$|\psi^{D}_{j}\rangle$\} is 
\begin{equation}\label{jjj}
    H^{1}=\left(\begin{array}{c c c c c c c c}
\Omega_{S} & 0 & 0 &0 & 0 &0&\frac{1}{\sqrt{2}}\Omega(t)&0\\
0& -\Omega_{S}& 0& 0& 0  &0&\frac{1}{\sqrt{2}}\Omega(t)&0\\
0 &0 & \Omega_{S}&0& 0& 0 &0 & \Omega(t) \\
0 &0 & 0&-\Omega_{S}& 0& 0 &0 & \Omega(t) \\
0 &0 & 0&0& \Omega(t)& 0 &0 & 0 \\
0 &0 & 0&0& 0& -\Omega(t) &0 & 0 \\
\frac{1}{\sqrt{2}}\Omega(t) &\frac{1}{\sqrt{2}}\Omega(t) & 0&0& 0& 0 &0 & 0 \\
0 &0 & \Omega(t)&\Omega(t)& 0& 0 &0 & 0 \\
\end{array}
\right).
\end{equation}
The corresponding eigenvalues are $E_{1}=\sqrt{\Omega(t)^{2}+\Omega_{S}^{2}},E_{2}=-\sqrt{\Omega(t)^{2}+\Omega_{S}^{2}},E_{3}=\sqrt{2\Omega(t)^{2}+\Omega_{S}^{2}},E_{4}=-\sqrt{2\Omega(t)^{2}+\Omega_{S}^{2}},E_{5}=\Omega(t),E_{6}=-\Omega(t),E_{7}=E_{8}=0$. Based on  Eq. ~(\ref{evolution}), one can find that  $\phi_{j}(4T)=\int_{0}^{4T}E_{j}dt$. Substituting the results in  Eq. ~(\ref{average1}), the corresponding average Hamiltonian of  Eq.~(\ref{initial222}) is obtained as 
\begin{eqnarray}\label{average2}
  \bar{H}&=&\frac{1}{4T}\int_{0}^{4T}\sqrt{\Omega(t)^{2}+\Omega_{S}^{2}} dt ( |\psi^{D}_{1}\rangle\langle \psi^{D}_{1}|-|\psi^{D}_{2}\rangle\langle \psi^{D}_{2}|)
  \nonumber\\
  &&+\frac{1}{4T}\int_{0}^{4T}\sqrt{2\Omega(t)^{2}+\Omega_{S}^{2}} dt (|\psi^{D}_{3}\rangle\langle \psi^{D}_{3}|- |\psi^{D}_{4}\rangle\langle\psi^{D}_{4}|)
  \nonumber\\
&&+\frac{1}{4T}\int_{0}^{4T}\Omega(t)dt(|\psi^{D}_{5}\rangle\langle \psi^{D}_{5}|- |\psi^{D}_{6}\rangle\langle\psi^{D}_{6}|)\nonumber\\
  &=&\frac{1}{4T}\int_{0}^{4T}\sqrt{\Omega(t)^{2}+\Omega_{S}^{2}} dt ( |\psi^{D}_{1}\rangle\langle \psi^{D}_{1}|-|\psi^{D}_{2}\rangle\langle \psi^{D}_{2}|)
  \nonumber\\
  &&+\frac{1}{4T}\int_{0}^{4T}\sqrt{2\Omega(t)^{2}+\Omega_{S}^{2}} dt (|\psi^{D}_{3}\rangle\langle \psi^{D}_{3}|- |\psi^{D}_{4}\rangle\langle\psi^{D}_{4}|)\nonumber\\
  &&
  +\frac{1}{4T}\int_{0}^{4T}\Omega_{m}\mathrm{exp}\big[-\frac{(t-2T)^{2}}{T^{2}}\big]dt\big[\frac{1}{\sqrt{2}}(|10\rangle-|01\rangle)\langle S_{0}|+\frac{1}{\sqrt{2}}|S_{0}\rangle(\langle 10|-\langle 01|)\big]\nonumber\\
&=&\frac{1}{4T}\int_{0}^{4T}\sqrt{\Omega(t)^{2}+\Omega_{S}^{2}} dt ( |\psi^{D}_{1}\rangle\langle \psi^{D}_{1}|-|\psi^{D}_{2}\rangle\langle \psi^{D}_{2}|)\
  \nonumber\\
&&+\frac{1}{4T}\int_{0}^{4T}\sqrt{2\Omega(t)^{2}+\Omega_{S}^{2}} dt (|\psi^{D}_{3}\rangle\langle \psi^{D}_{3}|- |\psi^{D}_{4}\rangle\langle\psi^{D}_{4}|)\nonumber\\
  &&
  +\frac{g}{\sqrt{2}}(|10\rangle-|01\rangle)\langle S_{0}|+\frac{g}{\sqrt{2}}|S_{0}\rangle(\langle 10|-\langle 01|),\nonumber
  \end{eqnarray}
where $g=\int_{0}^{4T}\Omega_{m}\mathrm{exp}[-\frac{(t-2T)^{2}}{T^{2}}]dt/(4T)=\sqrt{\pi}\Omega_{m}\mathrm{Erf}[2]/4$.
}

\bibliography{sn-article.bbl}


\begin{thebibliography}{105}
\ifx \bisbn   \undefined \def \bisbn  #1{ISBN #1}\fi
\ifx \binits  \undefined \def \binits#1{#1}\fi
\ifx \bauthor  \undefined \def \bauthor#1{#1}\fi
\ifx \batitle  \undefined \def \batitle#1{#1}\fi
\ifx \bjtitle  \undefined \def \bjtitle#1{#1}\fi
\ifx \bvolume  \undefined \def \bvolume#1{\textbf{#1}}\fi
\ifx \byear  \undefined \def \byear#1{#1}\fi
\ifx \bissue  \undefined \def \bissue#1{#1}\fi
\ifx \bfpage  \undefined \def \bfpage#1{#1}\fi
\ifx \blpage  \undefined \def \blpage #1{#1}\fi
\ifx \burl  \undefined \def \burl#1{\textsf{#1}}\fi
\ifx \doiurl  \undefined \def \doiurl#1{\url{https://doi.org/#1}}\fi
\ifx \betal  \undefined \def \betal{\textit{et al.}}\fi
\ifx \binstitute  \undefined \def \binstitute#1{#1}\fi
\ifx \binstitutionaled  \undefined \def \binstitutionaled#1{#1}\fi
\ifx \bctitle  \undefined \def \bctitle#1{#1}\fi
\ifx \beditor  \undefined \def \beditor#1{#1}\fi
\ifx \bpublisher  \undefined \def \bpublisher#1{#1}\fi
\ifx \bbtitle  \undefined \def \bbtitle#1{#1}\fi
\ifx \bedition  \undefined \def \bedition#1{#1}\fi
\ifx \bseriesno  \undefined \def \bseriesno#1{#1}\fi
\ifx \blocation  \undefined \def \blocation#1{#1}\fi
\ifx \bsertitle  \undefined \def \bsertitle#1{#1}\fi
\ifx \bsnm \undefined \def \bsnm#1{#1}\fi
\ifx \bsuffix \undefined \def \bsuffix#1{#1}\fi
\ifx \bparticle \undefined \def \bparticle#1{#1}\fi
\ifx \barticle \undefined \def \barticle#1{#1}\fi
\bibcommenthead
\ifx \bconfdate \undefined \def \bconfdate #1{#1}\fi
\ifx \botherref \undefined \def \botherref #1{#1}\fi
\ifx \url \undefined \def \url#1{\textsf{#1}}\fi
\ifx \bchapter \undefined \def \bchapter#1{#1}\fi
\ifx \bbook \undefined \def \bbook#1{#1}\fi
\ifx \bcomment \undefined \def \bcomment#1{#1}\fi
\ifx \oauthor \undefined \def \oauthor#1{#1}\fi
\ifx \citeauthoryear \undefined \def \citeauthoryear#1{#1}\fi
\ifx \endbibitem  \undefined \def \endbibitem {}\fi
\ifx \bconflocation  \undefined \def \bconflocation#1{#1}\fi
\ifx \arxivurl  \undefined \def \arxivurl#1{\textsf{#1}}\fi
\csname PreBibitemsHook\endcsname

\bibitem[\protect\citeauthoryear{Feynman}{1982}]{Feynman}
\begin{barticle}
\bauthor{\bsnm{Feynman}, \binits{R.P.}}:
\batitle{Simulating physics with computers}.
\bjtitle{Int. J. Theor. Phys.}
\bvolume{21}(\bissue{6}),
\bfpage{467}--\blpage{488}
(\byear{1982})
\doiurl{10.1007/BF02650179}
\end{barticle}
\endbibitem

\bibitem[\protect\citeauthoryear{Shor}{1997}]{shor}
\begin{barticle}
\bauthor{\bsnm{Shor}, \binits{P.W.}}:
\batitle{Polynomial-time algorithms for prime factorization and discrete
  logarithms on a quantum computer}.
\bjtitle{SIAM J. Comput.}
\bvolume{26}(\bissue{5}),
\bfpage{1484}--\blpage{1509}
(\byear{1997})
\doiurl{10.1137/S0097539795293172}
\end{barticle}
\endbibitem

\bibitem[\protect\citeauthoryear{Freedman et~al.}{2002}]{Freedman}
\begin{barticle}
\bauthor{\bsnm{Freedman}, \binits{M.H.}},
\bauthor{\bsnm{Kitaev}, \binits{A.}},
\bauthor{\bsnm{Wang}, \binits{Z.}}:
\batitle{Simulation of topological field theories by quantum computers}.
\bjtitle{Commun. Math. Phys.}
\bvolume{227}(\bissue{3}),
\bfpage{587}--\blpage{603}
(\byear{2002})
\doiurl{10.1007/s002200200635}
\end{barticle}
\endbibitem

\bibitem[\protect\citeauthoryear{Childs et~al.}{2003}]{Childs}
\begin{botherref}
\oauthor{\bsnm{Childs}, \binits{A.}},
\oauthor{\bsnm{Cleve}, \binits{R.}},
\oauthor{\bsnm{Deotto}, \binits{E.}},
\oauthor{\bsnm{Farhi}, \binits{E.}},
\oauthor{\bsnm{Gutmann}, \binits{S.}},
\oauthor{\bsnm{Spielman}, \binits{D.}}:
Proceedings of the 35th acm symposium on theory of computing (stoc 2003)
(2003)
\end{botherref}
\endbibitem

\bibitem[\protect\citeauthoryear{Hallgren}{2007}]{Hallgren}
\begin{botherref}
\oauthor{\bsnm{Hallgren}, \binits{S.}}:
Polynomial-time quantum algorithms for pell's equation and the principal ideal
  problem.
J. ACM
\textbf{54}(1)
(2007)
\doiurl{10.1145/1206035.1206039}
\end{botherref}
\endbibitem

\bibitem[\protect\citeauthoryear{Zanardi and Rasetti}{1999}]{Zanardi}
\begin{barticle}
\bauthor{\bsnm{Zanardi}, \binits{P.}},
\bauthor{\bsnm{Rasetti}, \binits{M.}}:
\batitle{Holonomic quantum computation}.
\bjtitle{Phys. Lett. A}
\bvolume{264}(\bissue{2}),
\bfpage{94}--\blpage{99}
(\byear{1999})
\doiurl{10.1016/S0375-9601(99)00803-8}
\end{barticle}
\endbibitem

\bibitem[\protect\citeauthoryear{Wilczek and Zee}{1984}]{Wilczek}
\begin{barticle}
\bauthor{\bsnm{Wilczek}, \binits{F.}},
\bauthor{\bsnm{Zee}, \binits{A.}}:
\batitle{Appearance of gauge structure in simple dynamical systems}.
\bjtitle{Phys. Rev. Lett.}
\bvolume{52},
\bfpage{2111}--\blpage{2114}
(\byear{1984})
\doiurl{10.1103/PhysRevLett.52.2111}
\end{barticle}
\endbibitem

\bibitem[\protect\citeauthoryear{Pachos et~al.}{1999}]{Pachos}
\begin{barticle}
\bauthor{\bsnm{Pachos}, \binits{J.}},
\bauthor{\bsnm{Zanardi}, \binits{P.}},
\bauthor{\bsnm{Rasetti}, \binits{M.}}:
\batitle{Non-abelian berry connections for quantum computation}.
\bjtitle{Phys. Rev. A}
\bvolume{61},
\bfpage{010305}
(\byear{1999})
\doiurl{10.1103/PhysRevA.61.010305}
\end{barticle}
\endbibitem

\bibitem[\protect\citeauthoryear{Duan et~al.}{2002}]{Duan}
\begin{barticle}
\bauthor{\bsnm{Duan}, \binits{L.-M.}},
\bauthor{\bsnm{Cirac}, \binits{J.I.}},
\bauthor{\bsnm{Zoller}, \binits{P.}}:
\batitle{Three-dimensional theory for interaction between atomic ensembles and
  free-space light}.
\bjtitle{Phys. Rev. A}
\bvolume{66},
\bfpage{023818}
(\byear{2002})
\doiurl{10.1103/PhysRevA.66.023818}
\end{barticle}
\endbibitem

\bibitem[\protect\citeauthoryear{Anandan}{1988}]{Anandan}
\begin{barticle}
\bauthor{\bsnm{Anandan}, \binits{J.}}:
\batitle{Non-adiabatic non-abelian geometric phase}.
\bjtitle{Phys. Lett. A}
\bvolume{133}(\bissue{4}),
\bfpage{171}--\blpage{175}
(\byear{1988})
\doiurl{10.1016/0375-9601(88)91010-9}
\end{barticle}
\endbibitem

\bibitem[\protect\citeauthoryear{Sjöqvist et~al.}{2012}]{Svist}
\begin{barticle}
\bauthor{\bsnm{Sjöqvist}, \binits{E.}},
\bauthor{\bsnm{Tong}, \binits{D.M.}},
\bauthor{\bsnm{Andersson}, \binits{L.M.}},
\bauthor{\bsnm{Hessmo}, \binits{B.}},
\bauthor{\bsnm{Johansson}, \binits{M.}},
\bauthor{\bsnm{Singh}, \binits{K.}}:
\batitle{Non-adiabatic holonomic quantum computation}.
\bjtitle{New J. Phys.}
\bvolume{14}(\bissue{10}),
\bfpage{103035}
(\byear{2012})
\doiurl{10.1088/1367-2630/14/10/103035}
\end{barticle}
\endbibitem

\bibitem[\protect\citeauthoryear{Xu et~al.}{2012}]{Xu}
\begin{barticle}
\bauthor{\bsnm{Xu}, \binits{G.F.}},
\bauthor{\bsnm{Zhang}, \binits{J.}},
\bauthor{\bsnm{Tong}, \binits{D.M.}},
\bauthor{\bsnm{Sj\"oqvist}, \binits{E.}},
\bauthor{\bsnm{Kwek}, \binits{L.C.}}:
\batitle{Nonadiabatic holonomic quantum computation in decoherence-free
  subspaces}.
\bjtitle{Phys. Rev. Lett.}
\bvolume{109},
\bfpage{170501}
(\byear{2012})
\doiurl{10.1103/PhysRevLett.109.170501}
\end{barticle}
\endbibitem

\bibitem[\protect\citeauthoryear{Mousolou et~al.}{2014}]{Mousolou}
\begin{barticle}
\bauthor{\bsnm{Mousolou}, \binits{V.A.}},
\bauthor{\bsnm{Canali}, \binits{C.M.}},
\bauthor{\bsnm{Sjöqvist}, \binits{E.}}:
\batitle{Universal non-adiabatic holonomic gates in quantum dots and
  single-molecule magnets}.
\bjtitle{New J. Phys.}
\bvolume{16}(\bissue{1}),
\bfpage{013029}
(\byear{2014})
\doiurl{10.1088/1367-2630/16/1/013029}
\end{barticle}
\endbibitem

\bibitem[\protect\citeauthoryear{Xu and Long}{2014}]{Xu3}
\begin{barticle}
\bauthor{\bsnm{Xu}, \binits{G.}},
\bauthor{\bsnm{Long}, \binits{G.}}:
\batitle{Protecting geometric gates by dynamical decoupling}.
\bjtitle{Phys. Rev. A}
\bvolume{90},
\bfpage{022323}
(\byear{2014})
\doiurl{10.1103/PhysRevA.90.022323}
\end{barticle}
\endbibitem

\bibitem[\protect\citeauthoryear{Xu et~al.}{2015}]{Xu1}
\begin{barticle}
\bauthor{\bsnm{Xu}, \binits{G.F.}},
\bauthor{\bsnm{Liu}, \binits{C.L.}},
\bauthor{\bsnm{Zhao}, \binits{P.Z.}},
\bauthor{\bsnm{Tong}, \binits{D.M.}}:
\batitle{Nonadiabatic holonomic gates realized by a single-shot
  implementation}.
\bjtitle{Phys. Rev. A}
\bvolume{92},
\bfpage{052302}
(\byear{2015})
\doiurl{10.1103/PhysRevA.92.052302}
\end{barticle}
\endbibitem

\bibitem[\protect\citeauthoryear{Herterich and Sj\"oqvist}{2016}]{Herterich}
\begin{barticle}
\bauthor{\bsnm{Herterich}, \binits{E.}},
\bauthor{\bsnm{Sj\"oqvist}, \binits{E.}}:
\batitle{Single-loop multiple-pulse nonadiabatic holonomic quantum gates}.
\bjtitle{Phys. Rev. A}
\bvolume{94},
\bfpage{052310}
(\byear{2016})
\doiurl{10.1103/PhysRevA.94.052310}
\end{barticle}
\endbibitem

\bibitem[\protect\citeauthoryear{Hong et~al.}{2018}]{Hong}
\begin{barticle}
\bauthor{\bsnm{Hong}, \binits{Z.-P.}},
\bauthor{\bsnm{Liu}, \binits{B.-J.}},
\bauthor{\bsnm{Cai}, \binits{J.-Q.}},
\bauthor{\bsnm{Zhang}, \binits{X.-D.}},
\bauthor{\bsnm{Hu}, \binits{Y.}},
\bauthor{\bsnm{Wang}, \binits{Z.D.}},
\bauthor{\bsnm{Xue}, \binits{Z.-Y.}}:
\batitle{Implementing universal nonadiabatic holonomic quantum gates with
  transmons}.
\bjtitle{Phys. Rev. A}
\bvolume{97},
\bfpage{022332}
(\byear{2018})
\doiurl{10.1103/PhysRevA.97.022332}
\end{barticle}
\endbibitem

\bibitem[\protect\citeauthoryear{Zhang et~al.}{2018}]{Zhang}
\begin{barticle}
\bauthor{\bsnm{Zhang}, \binits{J.}},
\bauthor{\bsnm{Devitt}, \binits{S.J.}},
\bauthor{\bsnm{You}, \binits{J.Q.}},
\bauthor{\bsnm{Nori}, \binits{F.}}:
\batitle{Holonomic surface codes for fault-tolerant quantum computation}.
\bjtitle{Phys. Rev. A}
\bvolume{97},
\bfpage{022335}
(\byear{2018})
\doiurl{10.1103/PhysRevA.97.022335}
\end{barticle}
\endbibitem

\bibitem[\protect\citeauthoryear{Xu et~al.}{2018}]{Xu2}
\begin{barticle}
\bauthor{\bsnm{Xu}, \binits{G.F.}},
\bauthor{\bsnm{Tong}, \binits{D.M.}},
\bauthor{\bsnm{Sj\"oqvist}, \binits{E.}}:
\batitle{Path-shortening realizations of nonadiabatic holonomic gates}.
\bjtitle{Phys. Rev. A}
\bvolume{98},
\bfpage{052315}
(\byear{2018})
\doiurl{10.1103/PhysRevA.98.052315}
\end{barticle}
\endbibitem

\bibitem[\protect\citeauthoryear{Liu et~al.}{2019}]{Liu}
\begin{barticle}
\bauthor{\bsnm{Liu}, \binits{B.-J.}},
\bauthor{\bsnm{Song}, \binits{X.-K.}},
\bauthor{\bsnm{Xue}, \binits{Z.-Y.}},
\bauthor{\bsnm{Wang}, \binits{X.}},
\bauthor{\bsnm{Yung}, \binits{M.-H.}}:
\batitle{Plug-and-play approach to nonadiabatic geometric quantum gates}.
\bjtitle{Phys. Rev. Lett.}
\bvolume{123},
\bfpage{100501}
(\byear{2019})
\doiurl{10.1103/PhysRevLett.123.100501}
\end{barticle}
\endbibitem

\bibitem[\protect\citeauthoryear{Chen et~al.}{2020}]{Chen}
\begin{barticle}
\bauthor{\bsnm{Chen}, \binits{T.}},
\bauthor{\bsnm{Shen}, \binits{P.}},
\bauthor{\bsnm{Xue}, \binits{Z.-Y.}}:
\batitle{Robust and fast holonomic quantum gates with encoding on
  superconducting circuits}.
\bjtitle{Phys. Rev. Appl.}
\bvolume{14},
\bfpage{034038}
(\byear{2020})
\doiurl{10.1103/PhysRevApplied.14.034038}
\end{barticle}
\endbibitem

\bibitem[\protect\citeauthoryear{Zhao et~al.}{2020}]{Zhao1}
\begin{barticle}
\bauthor{\bsnm{Zhao}, \binits{P.Z.}},
\bauthor{\bsnm{Li}, \binits{K.Z.}},
\bauthor{\bsnm{Xu}, \binits{G.F.}},
\bauthor{\bsnm{Tong}, \binits{D.M.}}:
\batitle{General approach for constructing hamiltonians for nonadiabatic
  holonomic quantum computation}.
\bjtitle{Phys. Rev. A}
\bvolume{101},
\bfpage{062306}
(\byear{2020})
\doiurl{10.1103/PhysRevA.101.062306}
\end{barticle}
\endbibitem

\bibitem[\protect\citeauthoryear{Wang et~al.}{2020}]{Wang}
\begin{barticle}
\bauthor{\bsnm{Wang}, \binits{Y.}},
\bauthor{\bsnm{Su}, \binits{Y.}},
\bauthor{\bsnm{Chen}, \binits{X.}},
\bauthor{\bsnm{Wu}, \binits{C.}}:
\batitle{Dephasing-protected scalable holonomic quantum computation on a rabi
  lattice}.
\bjtitle{Phys. Rev. Appl.}
\bvolume{14},
\bfpage{044043}
(\byear{2020})
\doiurl{10.1103/PhysRevApplied.14.044043}
\end{barticle}
\endbibitem

\bibitem[\protect\citeauthoryear{Liu et~al.}{2020}]{Liu1}
\begin{barticle}
\bauthor{\bsnm{Liu}, \binits{B.-J.}},
\bauthor{\bsnm{Su}, \binits{S.-L.}},
\bauthor{\bsnm{Yung}, \binits{M.-H.}}:
\batitle{Nonadiabatic noncyclic geometric quantum computation in rydberg
  atoms}.
\bjtitle{Phys. Rev. Res.}
\bvolume{2},
\bfpage{043130}
(\byear{2020})
\doiurl{10.1103/PhysRevResearch.2.043130}
\end{barticle}
\endbibitem

\bibitem[\protect\citeauthoryear{Shen et~al.}{2021}]{Shen}
\begin{barticle}
\bauthor{\bsnm{Shen}, \binits{P.}},
\bauthor{\bsnm{Chen}, \binits{T.}},
\bauthor{\bsnm{Xue}, \binits{Z.-Y.}}:
\batitle{Ultrafast holonomic quantum gates}.
\bjtitle{Phys. Rev. Appl.}
\bvolume{16},
\bfpage{044004}
(\byear{2021})
\doiurl{10.1103/PhysRevApplied.16.044004}
\end{barticle}
\endbibitem

\bibitem[\protect\citeauthoryear{Li and Xue}{2021}]{Li1}
\begin{barticle}
\bauthor{\bsnm{Li}, \binits{S.}},
\bauthor{\bsnm{Xue}, \binits{Z.-Y.}}:
\batitle{Dynamically corrected nonadiabatic holonomic quantum gates}.
\bjtitle{Phys. Rev. Appl.}
\bvolume{16},
\bfpage{044005}
(\byear{2021})
\doiurl{10.1103/PhysRevApplied.16.044005}
\end{barticle}
\endbibitem

\bibitem[\protect\citeauthoryear{Abdumalikov et~al.}{2013}]{Abdumalikov}
\begin{barticle}
\bauthor{\bsnm{Abdumalikov}, \binits{A.A.}},
\bauthor{\bsnm{Fink}, \binits{J.M.}},
\bauthor{\bsnm{Juliusson}, \binits{K.}},
\bauthor{\bsnm{Pechal}, \binits{M.}},
\bauthor{\bsnm{Berger}, \binits{S.}},
\bauthor{\bsnm{Wallraff}, \binits{A.}},
\bauthor{\bsnm{Filipp}, \binits{S.}}:
\batitle{Experimental realization of non-abelian non-adiabatic geometric
  gates}.
\bjtitle{Nature}
\bvolume{496}(\bissue{7446}),
\bfpage{482}--\blpage{485}
(\byear{2013})
\doiurl{10.1038/nature12010}
\end{barticle}
\endbibitem

\bibitem[\protect\citeauthoryear{Xu et~al.}{2018}]{XuY}
\begin{barticle}
\bauthor{\bsnm{Xu}, \binits{Y.}},
\bauthor{\bsnm{Cai}, \binits{W.}},
\bauthor{\bsnm{Ma}, \binits{Y.}},
\bauthor{\bsnm{Mu}, \binits{X.}},
\bauthor{\bsnm{Hu}, \binits{L.}},
\bauthor{\bsnm{Chen}, \binits{T.}},
\bauthor{\bsnm{Wang}, \binits{H.}},
\bauthor{\bsnm{Song}, \binits{Y.P.}},
\bauthor{\bsnm{Xue}, \binits{Z.-Y.}},
\bauthor{\bsnm{Yin}, \binits{Z.-q.}},
\bauthor{\bsnm{Sun}, \binits{L.}}:
\batitle{Single-loop realization of arbitrary nonadiabatic holonomic
  single-qubit quantum gates in a superconducting circuit}.
\bjtitle{Phys. Rev. Lett.}
\bvolume{121},
\bfpage{110501}
(\byear{2018})
\doiurl{10.1103/PhysRevLett.121.110501}
\end{barticle}
\endbibitem

\bibitem[\protect\citeauthoryear{Yan et~al.}{2019}]{YanT}
\begin{barticle}
\bauthor{\bsnm{Yan}, \binits{T.}},
\bauthor{\bsnm{Liu}, \binits{B.-J.}},
\bauthor{\bsnm{Xu}, \binits{K.}},
\bauthor{\bsnm{Song}, \binits{C.}},
\bauthor{\bsnm{Liu}, \binits{S.}},
\bauthor{\bsnm{Zhang}, \binits{Z.}},
\bauthor{\bsnm{Deng}, \binits{H.}},
\bauthor{\bsnm{Yan}, \binits{Z.}},
\bauthor{\bsnm{Rong}, \binits{H.}},
\bauthor{\bsnm{Huang}, \binits{K.}},
\bauthor{\bsnm{Yung}, \binits{M.-H.}},
\bauthor{\bsnm{Chen}, \binits{Y.}},
\bauthor{\bsnm{Yu}, \binits{D.}}:
\batitle{Experimental realization of nonadiabatic shortcut to non-abelian
  geometric gates}.
\bjtitle{Phys. Rev. Lett.}
\bvolume{122},
\bfpage{080501}
(\byear{2019})
\doiurl{10.1103/PhysRevLett.122.080501}
\end{barticle}
\endbibitem

\bibitem[\protect\citeauthoryear{Feng et~al.}{2013}]{Feng1}
\begin{barticle}
\bauthor{\bsnm{Feng}, \binits{G.}},
\bauthor{\bsnm{Xu}, \binits{G.}},
\bauthor{\bsnm{Long}, \binits{G.}}:
\batitle{Experimental realization of nonadiabatic holonomic quantum
  computation}.
\bjtitle{Phys. Rev. Lett.}
\bvolume{110},
\bfpage{190501}
(\byear{2013})
\doiurl{10.1103/PhysRevLett.110.190501}
\end{barticle}
\endbibitem

\bibitem[\protect\citeauthoryear{Li et~al.}{2017}]{LiH}
\begin{barticle}
\bauthor{\bsnm{Li}, \binits{H.}},
\bauthor{\bsnm{Liu}, \binits{Y.}},
\bauthor{\bsnm{Long}, \binits{G.}}:
\batitle{Experimental realization of single-shot nonadiabatic holonomic gates
  in nuclear spins}.
\bjtitle{Sci. China-Phys. Mech. Astron.}
\bvolume{60}(\bissue{8}),
\bfpage{080311}
(\byear{2017})
\doiurl{10.1007/s11433-017-9058-7}
\end{barticle}
\endbibitem

\bibitem[\protect\citeauthoryear{Zhu et~al.}{2019}]{ZhuZ}
\begin{barticle}
\bauthor{\bsnm{Zhu}, \binits{Z.}},
\bauthor{\bsnm{Chen}, \binits{T.}},
\bauthor{\bsnm{Yang}, \binits{X.}},
\bauthor{\bsnm{Bian}, \binits{J.}},
\bauthor{\bsnm{Xue}, \binits{Z.-Y.}},
\bauthor{\bsnm{Peng}, \binits{X.}}:
\batitle{Single-loop and composite-loop realization of nonadiabatic holonomic
  quantum gates in a decoherence-free subspace}.
\bjtitle{Phys. Rev. Appl.}
\bvolume{12},
\bfpage{024024}
(\byear{2019})
\doiurl{10.1103/PhysRevApplied.12.024024}
\end{barticle}
\endbibitem

\bibitem[\protect\citeauthoryear{Zu et~al.}{2014}]{ZuC}
\begin{barticle}
\bauthor{\bsnm{Zu}, \binits{C.}},
\bauthor{\bsnm{Wang}, \binits{W.-B.}},
\bauthor{\bsnm{He}, \binits{L.}},
\bauthor{\bsnm{Zhang}, \binits{W.-G.}},
\bauthor{\bsnm{Dai}, \binits{C.-Y.}},
\bauthor{\bsnm{Wang}, \binits{F.}},
\bauthor{\bsnm{Duan}, \binits{L.-M.}}:
\batitle{Experimental realization of universal geometric quantum gates with
  solid-state spins}.
\bjtitle{Nature}
\bvolume{514}(\bissue{7520}),
\bfpage{72}--\blpage{75}
(\byear{2014})
\doiurl{10.1038/nature13729}
\end{barticle}
\endbibitem

\bibitem[\protect\citeauthoryear{Arroyo-Camejo et~al.}{2014}]{Arroyo}
\begin{barticle}
\bauthor{\bsnm{Arroyo-Camejo}, \binits{S.}},
\bauthor{\bsnm{Lazariev}, \binits{A.}},
\bauthor{\bsnm{Hell}, \binits{S.W.}},
\bauthor{\bsnm{Balasubramanian}, \binits{G.}}:
\batitle{Room temperature high-fidelity holonomic single-qubit gate on a
  solid-state spin}.
\bjtitle{Nat. Commun.}
\bvolume{5}(\bissue{1}),
\bfpage{4870}
(\byear{2014})
\doiurl{10.1038/ncomms5870}
\end{barticle}
\endbibitem

\bibitem[\protect\citeauthoryear{Sekiguchi et~al.}{2017}]{Sekiguchi}
\begin{barticle}
\bauthor{\bsnm{Sekiguchi}, \binits{Y.}},
\bauthor{\bsnm{Niikura}, \binits{N.}},
\bauthor{\bsnm{Kuroiwa}, \binits{R.}},
\bauthor{\bsnm{Kano}, \binits{H.}},
\bauthor{\bsnm{Kosaka}, \binits{H.}}:
\batitle{Optical holonomic single quantum gates with a geometric spin under a
  zero field}.
\bjtitle{Nat. Photonics}
\bvolume{11}(\bissue{5}),
\bfpage{309}--\blpage{314}
(\byear{2017})
\doiurl{10.1038/nphoton.2017.40}
\end{barticle}
\endbibitem

\bibitem[\protect\citeauthoryear{Zhou et~al.}{2017}]{ZhouB}
\begin{barticle}
\bauthor{\bsnm{Zhou}, \binits{B.B.}},
\bauthor{\bsnm{Jerger}, \binits{P.C.}},
\bauthor{\bsnm{Shkolnikov}, \binits{V.O.}},
\bauthor{\bsnm{Heremans}, \binits{F.J.}},
\bauthor{\bsnm{Burkard}, \binits{G.}},
\bauthor{\bsnm{Awschalom}, \binits{D.D.}}:
\batitle{Holonomic quantum control by coherent optical excitation in diamond}.
\bjtitle{Phys. Rev. Lett.}
\bvolume{119},
\bfpage{140503}
(\byear{2017})
\doiurl{10.1103/PhysRevLett.119.140503}
\end{barticle}
\endbibitem

\bibitem[\protect\citeauthoryear{Nagata et~al.}{2018}]{Nagata}
\begin{barticle}
\bauthor{\bsnm{Nagata}, \binits{K.}},
\bauthor{\bsnm{Kuramitani}, \binits{K.}},
\bauthor{\bsnm{Sekiguchi}, \binits{Y.}},
\bauthor{\bsnm{Kosaka}, \binits{H.}}:
\batitle{Universal holonomic quantum gates over geometric spin qubits with
  polarised microwaves}.
\bjtitle{Nat. Commun.}
\bvolume{9}(\bissue{1}),
\bfpage{3227}
(\byear{2018})
\doiurl{10.1038/s41467-018-05664-w}
\end{barticle}
\endbibitem

\bibitem[\protect\citeauthoryear{Saffman et~al.}{2010}]{Saffman}
\begin{barticle}
\bauthor{\bsnm{Saffman}, \binits{M.}},
\bauthor{\bsnm{Walker}, \binits{T.G.}},
\bauthor{\bsnm{M\o{}lmer}, \binits{K.}}:
\batitle{Quantum information with rydberg atoms}.
\bjtitle{Rev. Mod. Phys.}
\bvolume{82},
\bfpage{2313}--\blpage{2363}
(\byear{2010})
\doiurl{10.1103/RevModPhys.82.2313}
\end{barticle}
\endbibitem

\bibitem[\protect\citeauthoryear{Browaeys et~al.}{2016}]{Browaeys}
\begin{barticle}
\bauthor{\bsnm{Browaeys}, \binits{A.}},
\bauthor{\bsnm{Barredo}, \binits{D.}},
\bauthor{\bsnm{Lahaye}, \binits{T.}}:
\batitle{Experimental investigations of dipole–dipole interactions between a
  few rydberg atoms}.
\bjtitle{J. Phys. B At. Mol. Opt. Phys.}
\bvolume{49}(\bissue{15}),
\bfpage{152001}
(\byear{2016})
\doiurl{10.1088/0953-4075/49/15/152001}
\end{barticle}
\endbibitem

\bibitem[\protect\citeauthoryear{Saffman}{2016}]{Saffman1}
\begin{barticle}
\bauthor{\bsnm{Saffman}, \binits{M.}}:
\batitle{Quantum computing with atomic qubits and rydberg interactions:
  progress and challenges}.
\bjtitle{J. Phys. B At. Mol. Opt. Phys.}
\bvolume{49}(\bissue{20}),
\bfpage{202001}
(\byear{2016})
\doiurl{10.1088/0953-4075/49/20/202001}
\end{barticle}
\endbibitem

\bibitem[\protect\citeauthoryear{Shao et~al.}{2023}]{ShaoXQ1}
\begin{barticle}
\bauthor{\bsnm{Shao}, \binits{X.Q.}},
\bauthor{\bsnm{Liu}, \binits{F.}},
\bauthor{\bsnm{Xue}, \binits{X.W.}},
\bauthor{\bsnm{Mu}, \binits{W.L.}},
\bauthor{\bsnm{Li}, \binits{W.}}:
\batitle{High-fidelity interconversion between greenberger-horne-zeilinger and
  $w$ states through floquet-lindblad engineering in rydberg atom arrays}.
\bjtitle{Phys. Rev. Appl.}
\bvolume{20},
\bfpage{014014}
(\byear{2023})
\doiurl{10.1103/PhysRevApplied.20.014014}
\end{barticle}
\endbibitem

\bibitem[\protect\citeauthoryear{Zhao et~al.}{2024}]{10.1063/5.0192602}
\begin{barticle}
\bauthor{\bsnm{Zhao}, \binits{Y.}},
\bauthor{\bsnm{Yang}, \binits{Y.-Q.}},
\bauthor{\bsnm{Li}, \binits{W.}},
\bauthor{\bsnm{Shao}, \binits{X.-Q.}}:
\batitle{{Dissipative stabilization of high-dimensional GHZ states for neutral
  atoms}}.
\bjtitle{Applied Physics Letters}
\bvolume{124}(\bissue{11}),
\bfpage{114001}
(\byear{2024})
\doiurl{10.1063/5.0192602}
\end{barticle}
\endbibitem

\bibitem[\protect\citeauthoryear{Jaksch et~al.}{2000}]{Jaksch}
\begin{barticle}
\bauthor{\bsnm{Jaksch}, \binits{D.}},
\bauthor{\bsnm{Cirac}, \binits{J.I.}},
\bauthor{\bsnm{Zoller}, \binits{P.}},
\bauthor{\bsnm{Rolston}, \binits{S.L.}},
\bauthor{\bsnm{C\^ot\'e}, \binits{R.}},
\bauthor{\bsnm{Lukin}, \binits{M.D.}}:
\batitle{Fast quantum gates for neutral atoms}.
\bjtitle{Phys. Rev. Lett.}
\bvolume{85},
\bfpage{2208}--\blpage{2211}
(\byear{2000})
\doiurl{10.1103/PhysRevLett.85.2208}
\end{barticle}
\endbibitem

\bibitem[\protect\citeauthoryear{Lukin et~al.}{2001}]{Lukin}
\begin{barticle}
\bauthor{\bsnm{Lukin}, \binits{M.D.}},
\bauthor{\bsnm{Fleischhauer}, \binits{M.}},
\bauthor{\bsnm{Cote}, \binits{R.}},
\bauthor{\bsnm{Duan}, \binits{L.M.}},
\bauthor{\bsnm{Jaksch}, \binits{D.}},
\bauthor{\bsnm{Cirac}, \binits{J.I.}},
\bauthor{\bsnm{Zoller}, \binits{P.}}:
\batitle{Dipole blockade and quantum information processing in mesoscopic
  atomic ensembles}.
\bjtitle{Phys. Rev. Lett.}
\bvolume{87},
\bfpage{037901}
(\byear{2001})
\doiurl{10.1103/PhysRevLett.87.037901}
\end{barticle}
\endbibitem

\bibitem[\protect\citeauthoryear{Urban et~al.}{2009}]{Urban}
\begin{barticle}
\bauthor{\bsnm{Urban}, \binits{E.}},
\bauthor{\bsnm{Johnson}, \binits{T.A.}},
\bauthor{\bsnm{Henage}, \binits{T.}},
\bauthor{\bsnm{Isenhower}, \binits{L.}},
\bauthor{\bsnm{Yavuz}, \binits{D.D.}},
\bauthor{\bsnm{Walker}, \binits{T.G.}},
\bauthor{\bsnm{Saffman}, \binits{M.}}:
\batitle{Observation of rydberg blockade between two atoms}.
\bjtitle{Nat. Phys.}
\bvolume{5}(\bissue{2}),
\bfpage{110}--\blpage{114}
(\byear{2009})
\doiurl{10.1038/nphys1178}
\end{barticle}
\endbibitem

\bibitem[\protect\citeauthoryear{Ga{\"e}tan et~al.}{2009}]{Gaetan}
\begin{barticle}
\bauthor{\bsnm{Ga{\"e}tan}, \binits{A.}},
\bauthor{\bsnm{Miroshnychenko}, \binits{Y.}},
\bauthor{\bsnm{Wilk}, \binits{T.}},
\bauthor{\bsnm{Chotia}, \binits{A.}},
\bauthor{\bsnm{Viteau}, \binits{M.}},
\bauthor{\bsnm{Comparat}, \binits{D.}},
\bauthor{\bsnm{Pillet}, \binits{P.}},
\bauthor{\bsnm{Browaeys}, \binits{A.}},
\bauthor{\bsnm{Grangier}, \binits{P.}}:
\batitle{Observation of collective excitation of two individual atoms in the
  rydberg blockade regime}.
\bjtitle{Nat. Phys.}
\bvolume{5}(\bissue{2}),
\bfpage{115}--\blpage{118}
(\byear{2009})
\doiurl{10.1038/nphys1183}
\end{barticle}
\endbibitem

\bibitem[\protect\citeauthoryear{Ates et~al.}{2007}]{Ates}
\begin{barticle}
\bauthor{\bsnm{Ates}, \binits{C.}},
\bauthor{\bsnm{Pohl}, \binits{T.}},
\bauthor{\bsnm{Pattard}, \binits{T.}},
\bauthor{\bsnm{Rost}, \binits{J.M.}}:
\batitle{Antiblockade in rydberg excitation of an ultracold lattice gas}.
\bjtitle{Phys. Rev. Lett.}
\bvolume{98},
\bfpage{023002}
(\byear{2007})
\doiurl{10.1103/PhysRevLett.98.023002}
\end{barticle}
\endbibitem

\bibitem[\protect\citeauthoryear{Amthor et~al.}{2010}]{Amthor}
\begin{barticle}
\bauthor{\bsnm{Amthor}, \binits{T.}},
\bauthor{\bsnm{Giese}, \binits{C.}},
\bauthor{\bsnm{Hofmann}, \binits{C.S.}},
\bauthor{\bsnm{Weidem\"uller}, \binits{M.}}:
\batitle{Evidence of antiblockade in an ultracold rydberg gas}.
\bjtitle{Phys. Rev. Lett.}
\bvolume{104},
\bfpage{013001}
(\byear{2010})
\doiurl{10.1103/PhysRevLett.104.013001}
\end{barticle}
\endbibitem

\bibitem[\protect\citeauthoryear{Su et~al.}{2016}]{SU2016}
\begin{barticle}
\bauthor{\bsnm{Su}, \binits{S.-L.}},
\bauthor{\bsnm{Liang}, \binits{E.}},
\bauthor{\bsnm{Zhang}, \binits{S.}},
\bauthor{\bsnm{Wen}, \binits{J.-J.}},
\bauthor{\bsnm{Sun}, \binits{L.-L.}},
\bauthor{\bsnm{Jin}, \binits{Z.}},
\bauthor{\bsnm{Zhu}, \binits{A.-D.}}:
\batitle{One-step implementation of the rydberg-rydberg-interaction gate}.
\bjtitle{Phys. Rev. A}
\bvolume{93},
\bfpage{012306}
(\byear{2016})
\doiurl{10.1103/PhysRevA.93.012306}
\end{barticle}
\endbibitem

\bibitem[\protect\citeauthoryear{Su et~al.}{2017}]{SU2017}
\begin{barticle}
\bauthor{\bsnm{Su}, \binits{S.-L.}},
\bauthor{\bsnm{Tian}, \binits{Y.}},
\bauthor{\bsnm{Shen}, \binits{H.Z.}},
\bauthor{\bsnm{Zang}, \binits{H.}},
\bauthor{\bsnm{Liang}, \binits{E.}},
\bauthor{\bsnm{Zhang}, \binits{S.}}:
\batitle{Applications of the modified rydberg antiblockade regime with
  simultaneous driving}.
\bjtitle{Phys. Rev. A}
\bvolume{96},
\bfpage{042335}
(\byear{2017})
\doiurl{10.1103/PhysRevA.96.042335}
\end{barticle}
\endbibitem

\bibitem[\protect\citeauthoryear{Su et~al.}{2018}]{SU2018}
\begin{barticle}
\bauthor{\bsnm{Su}, \binits{S.L.}},
\bauthor{\bsnm{Shen}, \binits{H.Z.}},
\bauthor{\bsnm{Liang}, \binits{E.}},
\bauthor{\bsnm{Zhang}, \binits{S.}}:
\batitle{One-step construction of the multiple-qubit rydberg controlled-phase
  gate}.
\bjtitle{Phys. Rev. A}
\bvolume{98},
\bfpage{032306}
(\byear{2018})
\doiurl{10.1103/PhysRevA.98.032306}
\end{barticle}
\endbibitem

\bibitem[\protect\citeauthoryear{Su et~al.}{2020a}]{SU202001}
\begin{barticle}
\bauthor{\bsnm{Su}, \binits{S.-L.}},
\bauthor{\bsnm{Guo}, \binits{F.-Q.}},
\bauthor{\bsnm{Tian}, \binits{L.}},
\bauthor{\bsnm{Zhu}, \binits{X.-Y.}},
\bauthor{\bsnm{Yan}, \binits{L.-L.}},
\bauthor{\bsnm{Liang}, \binits{E.-J.}},
\bauthor{\bsnm{Feng}, \binits{M.}}:
\batitle{Nondestructive rydberg parity meter and its applications}.
\bjtitle{Phys. Rev. A}
\bvolume{101},
\bfpage{012347}
(\byear{2020})
\doiurl{10.1103/PhysRevA.101.012347}
\end{barticle}
\endbibitem

\bibitem[\protect\citeauthoryear{Su et~al.}{2020b}]{Su_202009}
\begin{barticle}
\bauthor{\bsnm{Su}, \binits{S.-L.}},
\bauthor{\bsnm{Guo}, \binits{F.-Q.}},
\bauthor{\bsnm{Wu}, \binits{J.-L.}},
\bauthor{\bsnm{Jin}, \binits{Z.}},
\bauthor{\bsnm{Shao}, \binits{X.Q.}},
\bauthor{\bsnm{Zhang}, \binits{S.}}:
\batitle{Rydberg antiblockade regimes: Dynamics and applications}.
\bjtitle{Europhysics Letters}
\bvolume{131}(\bissue{5}),
\bfpage{53001}
(\byear{2020})
\doiurl{10.1209/0295-5075/131/53001}
\end{barticle}
\endbibitem

\bibitem[\protect\citeauthoryear{M\o{}ller et~al.}{2008}]{Mller}
\begin{barticle}
\bauthor{\bsnm{M\o{}ller}, \binits{D.}},
\bauthor{\bsnm{Madsen}, \binits{L.B.}},
\bauthor{\bsnm{M\o{}lmer}, \binits{K.}}:
\batitle{Quantum gates and multiparticle entanglement by rydberg excitation
  blockade and adiabatic passage}.
\bjtitle{Phys. Rev. Lett.}
\bvolume{100},
\bfpage{170504}
(\byear{2008})
\doiurl{10.1103/PhysRevLett.100.170504}
\end{barticle}
\endbibitem

\bibitem[\protect\citeauthoryear{Carr and Saffman}{2013}]{Carr}
\begin{barticle}
\bauthor{\bsnm{Carr}, \binits{A.W.}},
\bauthor{\bsnm{Saffman}, \binits{M.}}:
\batitle{Preparation of entangled and antiferromagnetic states by dissipative
  rydberg pumping}.
\bjtitle{Phys. Rev. Lett.}
\bvolume{111},
\bfpage{033607}
(\byear{2013})
\doiurl{10.1103/PhysRevLett.111.033607}
\end{barticle}
\endbibitem

\bibitem[\protect\citeauthoryear{Tian et~al.}{2015}]{Tian}
\begin{barticle}
\bauthor{\bsnm{Tian}, \binits{X.-D.}},
\bauthor{\bsnm{Liu}, \binits{Y.-M.}},
\bauthor{\bsnm{Cui}, \binits{C.-L.}},
\bauthor{\bsnm{Wu}, \binits{J.-H.}}:
\batitle{Population transfer and quantum entanglement implemented in cold atoms
  involving two rydberg states via an adiabatic passage}.
\bjtitle{Phys. Rev. A}
\bvolume{92},
\bfpage{063411}
(\byear{2015})
\doiurl{10.1103/PhysRevA.92.063411}
\end{barticle}
\endbibitem

\bibitem[\protect\citeauthoryear{Su et~al.}{2015}]{Su}
\begin{barticle}
\bauthor{\bsnm{Su}, \binits{S.-L.}},
\bauthor{\bsnm{Guo}, \binits{Q.}},
\bauthor{\bsnm{Wang}, \binits{H.-F.}},
\bauthor{\bsnm{Zhang}, \binits{S.}}:
\batitle{Simplified scheme for entanglement preparation with rydberg pumping
  via dissipation}.
\bjtitle{Phys. Rev. A}
\bvolume{92},
\bfpage{022328}
(\byear{2015})
\doiurl{10.1103/PhysRevA.92.022328}
\end{barticle}
\endbibitem

\bibitem[\protect\citeauthoryear{Shao et~al.}{2017}]{Shao}
\begin{barticle}
\bauthor{\bsnm{Shao}, \binits{X.Q.}},
\bauthor{\bsnm{Wu}, \binits{J.H.}},
\bauthor{\bsnm{Yi}, \binits{X.X.}}:
\batitle{Dissipation-based entanglement via quantum zeno dynamics and rydberg
  antiblockade}.
\bjtitle{Phys. Rev. A}
\bvolume{95},
\bfpage{062339}
(\byear{2017})
\doiurl{10.1103/PhysRevA.95.062339}
\end{barticle}
\endbibitem

\bibitem[\protect\citeauthoryear{Zeng et~al.}{2017}]{Zeng}
\begin{barticle}
\bauthor{\bsnm{Zeng}, \binits{Y.}},
\bauthor{\bsnm{Xu}, \binits{P.}},
\bauthor{\bsnm{He}, \binits{X.}},
\bauthor{\bsnm{Liu}, \binits{Y.}},
\bauthor{\bsnm{Liu}, \binits{M.}},
\bauthor{\bsnm{Wang}, \binits{J.}},
\bauthor{\bsnm{Papoular}, \binits{D.J.}},
\bauthor{\bsnm{Shlyapnikov}, \binits{G.V.}},
\bauthor{\bsnm{Zhan}, \binits{M.}}:
\batitle{Entangling two individual atoms of different isotopes via rydberg
  blockade}.
\bjtitle{Phys. Rev. Lett.}
\bvolume{119},
\bfpage{160502}
(\byear{2017})
\doiurl{10.1103/PhysRevLett.119.160502}
\end{barticle}
\endbibitem

\bibitem[\protect\citeauthoryear{Shi}{2018}]{Shi}
\begin{barticle}
\bauthor{\bsnm{Shi}, \binits{X.-F.}}:
\batitle{Universal barenco quantum gates via a tunable noncollinear
  interaction}.
\bjtitle{Phys. Rev. A}
\bvolume{97},
\bfpage{032310}
(\byear{2018})
\doiurl{10.1103/PhysRevA.97.032310}
\end{barticle}
\endbibitem

\bibitem[\protect\citeauthoryear{Petrosyan and M\o{}lmer}{2018}]{Petrosyan}
\begin{barticle}
\bauthor{\bsnm{Petrosyan}, \binits{D.}},
\bauthor{\bsnm{M\o{}lmer}, \binits{K.}}:
\batitle{Deterministic free-space source of single photons using rydberg
  atoms}.
\bjtitle{Phys. Rev. Lett.}
\bvolume{121},
\bfpage{123605}
(\byear{2018})
\doiurl{10.1103/PhysRevLett.121.123605}
\end{barticle}
\endbibitem

\bibitem[\protect\citeauthoryear{Li and Shao}{2018}]{LDX}
\begin{barticle}
\bauthor{\bsnm{Li}, \binits{D.X.}},
\bauthor{\bsnm{Shao}, \binits{X.Q.}}:
\batitle{Unconventional rydberg pumping and applications in quantum information
  processing}.
\bjtitle{Phys. Rev. A}
\bvolume{98},
\bfpage{062338}
(\byear{2018})
\doiurl{10.1103/PhysRevA.98.062338}
\end{barticle}
\endbibitem

\bibitem[\protect\citeauthoryear{Omran et~al.}{2019}]{Omran}
\begin{barticle}
\bauthor{\bsnm{Omran}, \binits{A.}},
\bauthor{\bsnm{Levine}, \binits{H.}},
\bauthor{\bsnm{Keesling}, \binits{A.}},
\bauthor{\bsnm{Semeghini}, \binits{G.}},
\bauthor{\bsnm{Wang}, \binits{T.T.}},
\bauthor{\bsnm{Ebadi}, \binits{S.}},
\bauthor{\bsnm{Bernien}, \binits{H.}},
\bauthor{\bsnm{Zibrov}, \binits{A.S.}},
\bauthor{\bsnm{Pichler}, \binits{H.}},
\bauthor{\bsnm{Choi}, \binits{S.}},
\bauthor{\bsnm{Cui}, \binits{J.}},
\bauthor{\bsnm{Rossignolo}, \binits{M.}},
\bauthor{\bsnm{Rembold}, \binits{P.}},
\bauthor{\bsnm{Montangero}, \binits{S.}},
\bauthor{\bsnm{Calarco}, \binits{T.}},
\bauthor{\bsnm{Endres}, \binits{M.}},
\bauthor{\bsnm{Greiner}, \binits{M.}},
\bauthor{\bsnm{Vuletić}, \binits{V.}},
\bauthor{\bsnm{Lukin}, \binits{M.D.}}:
\batitle{Generation and manipulation of schr\"{o}dinger cat states in rydberg
  atom arrays}.
\bjtitle{Science}
\bvolume{365}(\bissue{6453}),
\bfpage{570}--\blpage{574}
(\byear{2019})
\doiurl{10.1126/science.aax9743}
\end{barticle}
\endbibitem

\bibitem[\protect\citeauthoryear{Wintermantel et~al.}{2020}]{Wintermantel}
\begin{barticle}
\bauthor{\bsnm{Wintermantel}, \binits{T.M.}},
\bauthor{\bsnm{Wang}, \binits{Y.}},
\bauthor{\bsnm{Lochead}, \binits{G.}},
\bauthor{\bsnm{Shevate}, \binits{S.}},
\bauthor{\bsnm{Brennen}, \binits{G.K.}},
\bauthor{\bsnm{Whitlock}, \binits{S.}}:
\batitle{Unitary and nonunitary quantum cellular automata with rydberg arrays}.
\bjtitle{Phys. Rev. Lett.}
\bvolume{124},
\bfpage{070503}
(\byear{2020})
\doiurl{10.1103/PhysRevLett.124.070503}
\end{barticle}
\endbibitem

\bibitem[\protect\citeauthoryear{Su et~al.}{2020}]{Su1}
\begin{barticle}
\bauthor{\bsnm{Su}, \binits{S.-L.}},
\bauthor{\bsnm{Guo}, \binits{F.-Q.}},
\bauthor{\bsnm{Tian}, \binits{L.}},
\bauthor{\bsnm{Zhu}, \binits{X.-Y.}},
\bauthor{\bsnm{Yan}, \binits{L.-L.}},
\bauthor{\bsnm{Liang}, \binits{E.-J.}},
\bauthor{\bsnm{Feng}, \binits{M.}}:
\batitle{Nondestructive rydberg parity meter and its applications}.
\bjtitle{Phys. Rev. A}
\bvolume{101},
\bfpage{012347}
(\byear{2020})
\doiurl{10.1103/PhysRevA.101.012347}
\end{barticle}
\endbibitem

\bibitem[\protect\citeauthoryear{Bai et~al.}{2020}]{Bai}
\begin{barticle}
\bauthor{\bsnm{Bai}, \binits{S.}},
\bauthor{\bsnm{Tian}, \binits{X.}},
\bauthor{\bsnm{Han}, \binits{X.}},
\bauthor{\bsnm{Jiao}, \binits{Y.}},
\bauthor{\bsnm{Wu}, \binits{J.}},
\bauthor{\bsnm{Zhao}, \binits{J.}},
\bauthor{\bsnm{Jia}, \binits{S.}}:
\batitle{Distinct antiblockade features of strongly interacting rydberg atoms
  under a two-color weak excitation scheme}.
\bjtitle{New J. Phys.}
\bvolume{22}(\bissue{1}),
\bfpage{013004}
(\byear{2020})
\doiurl{10.1088/1367-2630/ab6575}
\end{barticle}
\endbibitem

\bibitem[\protect\citeauthoryear{Yin et~al.}{2020}]{Yin:20}
\begin{barticle}
\bauthor{\bsnm{Yin}, \binits{H.-D.}},
\bauthor{\bsnm{Li}, \binits{X.-X.}},
\bauthor{\bsnm{Wang}, \binits{G.-C.}},
\bauthor{\bsnm{Shao}, \binits{X.-Q.}}:
\batitle{One-step implementation of toffoli gate for neutral atoms based on
  unconventional rydberg pumping}.
\bjtitle{Opt. Express}
\bvolume{28}(\bissue{24}),
\bfpage{35576}--\blpage{35587}
(\byear{2020})
\doiurl{10.1364/OE.410158}
\end{barticle}
\endbibitem

\bibitem[\protect\citeauthoryear{Yin and Shao}{2021}]{Yin:21}
\begin{barticle}
\bauthor{\bsnm{Yin}, \binits{H.-D.}},
\bauthor{\bsnm{Shao}, \binits{X.-Q.}}:
\batitle{Gaussian soft control-based quantum fan-out gate in ground-state
  manifolds of neutral atoms}.
\bjtitle{Opt. Lett.}
\bvolume{46}(\bissue{10}),
\bfpage{2541}--\blpage{2544}
(\byear{2021})
\doiurl{10.1364/OL.424469}
\end{barticle}
\endbibitem

\bibitem[\protect\citeauthoryear{Wu et~al.}{2021}]{Wu:21}
\begin{barticle}
\bauthor{\bsnm{Wu}, \binits{J.-L.}},
\bauthor{\bsnm{Wang}, \binits{Y.}},
\bauthor{\bsnm{Han}, \binits{J.-X.}},
\bauthor{\bsnm{Feng}, \binits{Y.-K.}},
\bauthor{\bsnm{Su}, \binits{S.-L.}},
\bauthor{\bsnm{Xia}, \binits{Y.}},
\bauthor{\bsnm{Jiang}, \binits{Y.}},
\bauthor{\bsnm{Song}, \binits{J.}}:
\batitle{One-step implementation of rydberg-antiblockade swap and
  controlled-swap gates with modified robustness}.
\bjtitle{Photon. Res.}
\bvolume{9}(\bissue{5}),
\bfpage{814}--\blpage{821}
(\byear{2021})
\doiurl{10.1364/PRJ.415795}
\end{barticle}
\endbibitem

\bibitem[\protect\citeauthoryear{Shi and Lu}{2021}]{SHI2021007}
\begin{barticle}
\bauthor{\bsnm{Shi}, \binits{X.-F.}},
\bauthor{\bsnm{Lu}, \binits{Y.}}:
\batitle{Quantum gates with weak van der waals interactions of neutral rydberg
  atoms}.
\bjtitle{Phys. Rev. A}
\bvolume{104},
\bfpage{012615}
(\byear{2021})
\doiurl{10.1103/PhysRevA.104.012615}
\end{barticle}
\endbibitem

\bibitem[\protect\citeauthoryear{Shi}{2021}]{SHI202110}
\begin{barticle}
\bauthor{\bsnm{Shi}, \binits{X.-F.}}:
\batitle{Hyperentanglement of divalent neutral atoms by rydberg blockade}.
\bjtitle{Phys. Rev. A}
\bvolume{104},
\bfpage{042422}
(\byear{2021})
\doiurl{10.1103/PhysRevA.104.042422}
\end{barticle}
\endbibitem

\bibitem[\protect\citeauthoryear{Li et~al.}{2022}]{PhysRevApplied.18.044042}
\begin{barticle}
\bauthor{\bsnm{Li}, \binits{X.X.}},
\bauthor{\bsnm{Shao}, \binits{X.Q.}},
\bauthor{\bsnm{Li}, \binits{W.}}:
\batitle{Single temporal-pulse-modulated parameterized controlled-phase gate
  for rydberg atoms}.
\bjtitle{Phys. Rev. Appl.}
\bvolume{18},
\bfpage{044042}
(\byear{2022})
\doiurl{10.1103/PhysRevApplied.18.044042}
\end{barticle}
\endbibitem

\bibitem[\protect\citeauthoryear{Shi}{2022}]{Shi_2022}
\begin{barticle}
\bauthor{\bsnm{Shi}, \binits{X.-F.}}:
\batitle{Quantum logic and entanglement by neutral rydberg atoms: methods and
  fidelity}.
\bjtitle{Quantum Science and Technology}
\bvolume{7}(\bissue{2}),
\bfpage{023002}
(\byear{2022})
\doiurl{10.1088/2058-9565/ac18b8}
\end{barticle}
\endbibitem

\bibitem[\protect\citeauthoryear{Jandura and
  Pupillo}{2022}]{Jandura2022timeoptimaltwothree}
\begin{barticle}
\bauthor{\bsnm{Jandura}, \binits{S.}},
\bauthor{\bsnm{Pupillo}, \binits{G.}}:
\batitle{Time-{O}ptimal {T}wo- and {T}hree-{Q}ubit {G}ates for {R}ydberg
  {A}toms}.
\bjtitle{{Quantum}}
\bvolume{6},
\bfpage{712}
(\byear{2022})
\doiurl{10.22331/q-2022-05-13-712}
\end{barticle}
\endbibitem

\bibitem[\protect\citeauthoryear{Evered et~al.}{2023}]{Evered2023}
\begin{barticle}
\bauthor{\bsnm{Evered}, \binits{S.J.}},
\bauthor{\bsnm{Bluvstein}, \binits{D.}},
\bauthor{\bsnm{Kalinowski}, \binits{M.}},
\bauthor{\bsnm{Ebadi}, \binits{S.}},
\bauthor{\bsnm{Manovitz}, \binits{T.}},
\bauthor{\bsnm{Zhou}, \binits{H.}},
\bauthor{\bsnm{Li}, \binits{S.H.}},
\bauthor{\bsnm{Geim}, \binits{A.A.}},
\bauthor{\bsnm{Wang}, \binits{T.T.}},
\bauthor{\bsnm{Maskara}, \binits{N.}},
\bauthor{\bsnm{Levine}, \binits{H.}},
\bauthor{\bsnm{Semeghini}, \binits{G.}},
\bauthor{\bsnm{Greiner}, \binits{M.}},
\bauthor{\bsnm{Vuleti{\'{c}}}, \binits{V.}},
\bauthor{\bsnm{Lukin}, \binits{M.D.}}:
\batitle{High-fidelity parallel entangling gates on a neutral-atom quantum
  computer}.
\bjtitle{Nature}
\bvolume{622}(\bissue{7982}),
\bfpage{268}--\blpage{272}
(\byear{2023})
\doiurl{10.1038/s41586-023-06481-y}
\end{barticle}
\endbibitem

\bibitem[\protect\citeauthoryear{Shi}{2023a}]{SHI202302}
\begin{barticle}
\bauthor{\bsnm{Shi}, \binits{X.-F.}}:
\batitle{Coherence-preserving cooling of nuclear-spin qubits in a weak magnetic
  field}.
\bjtitle{Phys. Rev. A}
\bvolume{107},
\bfpage{023102}
(\byear{2023})
\doiurl{10.1103/PhysRevA.107.023102}
\end{barticle}
\endbibitem

\bibitem[\protect\citeauthoryear{Shi}{2023b}]{Shi202310}
\begin{barticle}
\bauthor{\bsnm{Shi}, \binits{X.-F.}}:
\batitle{Fast nuclear-spin gates and electrons-nuclei entanglement of neutral
  atoms in weak magnetic fields}.
\bjtitle{Frontiers of Physics}
\bvolume{19}(\bissue{2}),
\bfpage{22203}
(\byear{2023})
\doiurl{10.1007/s11467-023-1332-0}
\end{barticle}
\endbibitem

\bibitem[\protect\citeauthoryear{Li et~al.}{2024}]{PhysRevA.109.042604}
\begin{barticle}
\bauthor{\bsnm{Li}, \binits{X.X.}},
\bauthor{\bsnm{Li}, \binits{D.X.}},
\bauthor{\bsnm{Shao}, \binits{X.Q.}}:
\batitle{Generation of complete graph states in a spin-1/2 heisenberg chain
  with a globally optimized magnetic field}.
\bjtitle{Phys. Rev. A}
\bvolume{109},
\bfpage{042604}
(\byear{2024})
\doiurl{10.1103/PhysRevA.109.042604}
\end{barticle}
\endbibitem

\bibitem[\protect\citeauthoryear{Shao}{2020}]{ShaoXQ}
\begin{barticle}
\bauthor{\bsnm{Shao}, \binits{X.-Q.}}:
\batitle{Selective rydberg pumping via strong dipole blockade}.
\bjtitle{Phys. Rev. A}
\bvolume{102},
\bfpage{053118}
(\byear{2020})
\doiurl{10.1103/PhysRevA.102.053118}
\end{barticle}
\endbibitem

\bibitem[\protect\citeauthoryear{Haase et~al.}{2018}]{Haase}
\begin{barticle}
\bauthor{\bsnm{Haase}, \binits{J.F.}},
\bauthor{\bsnm{Wang}, \binits{Z.-Y.}},
\bauthor{\bsnm{Casanova}, \binits{J.}},
\bauthor{\bsnm{Plenio}, \binits{M.B.}}:
\batitle{Soft quantum control for highly selective interactions among joint
  quantum systems}.
\bjtitle{Phys. Rev. Lett.}
\bvolume{121},
\bfpage{050402}
(\byear{2018})
\doiurl{10.1103/PhysRevLett.121.050402}
\end{barticle}
\endbibitem

\bibitem[\protect\citeauthoryear{Han et~al.}{2021}]{Han1}
\begin{barticle}
\bauthor{\bsnm{Han}, \binits{J.-X.}},
\bauthor{\bsnm{Wu}, \binits{J.-L.}},
\bauthor{\bsnm{Wang}, \binits{Y.}},
\bauthor{\bsnm{Xia}, \binits{Y.}},
\bauthor{\bsnm{Jiang}, \binits{Y.-Y.}},
\bauthor{\bsnm{Song}, \binits{J.}}:
\batitle{Large-scale greenberger-horne-zeilinger states through a topologically
  protected zero-energy mode in a superconducting qutrit-resonator chain}.
\bjtitle{Phys. Rev. A}
\bvolume{103},
\bfpage{032402}
(\byear{2021})
\doiurl{10.1103/PhysRevA.103.032402}
\end{barticle}
\endbibitem

\bibitem[\protect\citeauthoryear{Ravets et~al.}{2014}]{Sylvain}
\begin{barticle}
\bauthor{\bsnm{Ravets}, \binits{S.}},
\bauthor{\bsnm{Labuhn}, \binits{H.}},
\bauthor{\bsnm{Barredo}, \binits{D.}},
\bauthor{\bsnm{Béguin}, \binits{L.}},
\bauthor{\bsnm{Lahaye}, \binits{T.}},
\bauthor{\bsnm{Browaeys}, \binits{A.}}:
\batitle{Coherent dipole–dipole coupling between two single rydberg atoms at
  an electrically-tuned förster resonance}.
\bjtitle{Nat. Phys.}
\bvolume{10}(\bissue{12}),
\bfpage{914}--\blpage{917}
(\byear{2014})
\doiurl{10.1038/nphys3119}
\end{barticle}
\endbibitem

\bibitem[\protect\citeauthoryear{Ravets et~al.}{2015}]{PhysRevA.92.020701}
\begin{barticle}
\bauthor{\bsnm{Ravets}, \binits{S.}},
\bauthor{\bsnm{Labuhn}, \binits{H.}},
\bauthor{\bsnm{Barredo}, \binits{D.}},
\bauthor{\bsnm{Lahaye}, \binits{T.}},
\bauthor{\bsnm{Browaeys}, \binits{A.}}:
\batitle{Measurement of the angular dependence of the dipole-dipole interaction
  between two individual rydberg atoms at a f\"orster resonance}.
\bjtitle{Phys. Rev. A}
\bvolume{92},
\bfpage{020701}
(\byear{2015})
\doiurl{10.1103/PhysRevA.92.020701}
\end{barticle}
\endbibitem

\bibitem[\protect\citeauthoryear{Ashkarin
  et~al.}{2023}]{ashkarin2023highfidelity}
\begin{botherref}
\oauthor{\bsnm{Ashkarin}, \binits{I.}},
\oauthor{\bsnm{Lepoutre}, \binits{S.}},
\oauthor{\bsnm{Pillet}, \binits{P.}},
\oauthor{\bsnm{Beterov}, \binits{I.}},
\oauthor{\bsnm{Ryabtsev}, \binits{I.}},
\oauthor{\bsnm{Cheinet}, \binits{P.}}:
High-fidelity $CCR_Z(\phi)$ gates via RF-induced F\"{o}rster resonances
(2023).
\url{https://doi.org/10.48550/arXiv.2307.12789}
\end{botherref}
\endbibitem

\bibitem[\protect\citeauthoryear{Wang and Plenio}{2016}]{Wangzy}
\begin{barticle}
\bauthor{\bsnm{Wang}, \binits{Z.-Y.}},
\bauthor{\bsnm{Plenio}, \binits{M.B.}}:
\batitle{Necessary and sufficient condition for quantum adiabatic evolution by
  unitary control fields}.
\bjtitle{Phys. Rev. A}
\bvolume{93},
\bfpage{052107}
(\byear{2016})
\doiurl{10.1103/PhysRevA.93.052107}
\end{barticle}
\endbibitem

\bibitem[\protect\citeauthoryear{Xu et~al.}{2019}]{Xuk}
\begin{barticle}
\bauthor{\bsnm{Xu}, \binits{K.}},
\bauthor{\bsnm{Xie}, \binits{T.}},
\bauthor{\bsnm{Shi}, \binits{F.}},
\bauthor{\bsnm{Wang}, \binits{Z.-Y.}},
\bauthor{\bsnm{Xu}, \binits{X.}},
\bauthor{\bsnm{Wang}, \binits{P.}},
\bauthor{\bsnm{Wang}, \binits{Y.}},
\bauthor{\bsnm{Plenio}, \binits{M.B.}},
\bauthor{\bsnm{Du}, \binits{J.}}:
\batitle{Breaking the quantum adiabatic speed limit by jumping along
  geodesics}.
\bjtitle{Sci. Adv.}
\bvolume{5}(\bissue{6}),
\bfpage{3800}
(\byear{2019})
\doiurl{10.1126/sciadv.aax3800}
\end{barticle}
\endbibitem

\bibitem[\protect\citeauthoryear{Griffiths}{2005}]{Griffiths}
\begin{bbook}
\beditor{\bsnm{Griffiths}, \binits{D.J.}} (ed.):
\bbtitle{Introduction to Quantum Mechanics}.
\bpublisher{Pearson Prentice-Hall},
\blocation{New Jersey}
(\byear{2005})
\end{bbook}
\endbibitem

\bibitem[\protect\citeauthoryear{Levine et~al.}{2019}]{Levine}
\begin{barticle}
\bauthor{\bsnm{Levine}, \binits{H.}},
\bauthor{\bsnm{Keesling}, \binits{A.}},
\bauthor{\bsnm{Semeghini}, \binits{G.}},
\bauthor{\bsnm{Omran}, \binits{A.}},
\bauthor{\bsnm{Wang}, \binits{T.T.}},
\bauthor{\bsnm{Ebadi}, \binits{S.}},
\bauthor{\bsnm{Bernien}, \binits{H.}},
\bauthor{\bsnm{Greiner}, \binits{M.}},
\bauthor{\bsnm{Vuleti\'c}, \binits{V.}},
\bauthor{\bsnm{Pichler}, \binits{H.}},
\bauthor{\bsnm{Lukin}, \binits{M.D.}}:
\batitle{Parallel implementation of high-fidelity multiqubit gates with neutral
  atoms}.
\bjtitle{Phys. Rev. Lett.}
\bvolume{123},
\bfpage{170503}
(\byear{2019})
\doiurl{10.1103/PhysRevLett.123.170503}
\end{barticle}
\endbibitem

\bibitem[\protect\citeauthoryear{Madjarov et~al.}{2020}]{Madjarov}
\begin{barticle}
\bauthor{\bsnm{Madjarov}, \binits{I.S.}},
\bauthor{\bsnm{Covey}, \binits{J.P.}},
\bauthor{\bsnm{Shaw}, \binits{A.L.}},
\bauthor{\bsnm{Choi}, \binits{J.}},
\bauthor{\bsnm{Kale}, \binits{A.}},
\bauthor{\bsnm{Cooper}, \binits{A.}},
\bauthor{\bsnm{Pichler}, \binits{H.}},
\bauthor{\bsnm{Schkolnik}, \binits{V.}},
\bauthor{\bsnm{Williams}, \binits{J.R.}},
\bauthor{\bsnm{Endres}, \binits{M.}}:
\batitle{High-fidelity entanglement and detection of alkaline-earth rydberg
  atoms}.
\bjtitle{Nat. Phys.}
\bvolume{16}(\bissue{8}),
\bfpage{857}--\blpage{861}
(\byear{2020})
\doiurl{10.1038/s41567-020-0903-z}
\end{barticle}
\endbibitem

\bibitem[\protect\citeauthoryear{Ozaydin et~al.}{2014}]{Ozaydin}
\begin{barticle}
\bauthor{\bsnm{Ozaydin}, \binits{F.}},
\bauthor{\bsnm{Bugu}, \binits{S.}},
\bauthor{\bsnm{Yesilyurt}, \binits{C.}},
\bauthor{\bsnm{Altintas}, \binits{A.A.}},
\bauthor{\bsnm{Tame}, \binits{M.}},
\bauthor{\bsnm{\"Ozdemir}, \binits{S.K.}}:
\batitle{Fusing multiple $w$ states simultaneously with a fredkin gate}.
\bjtitle{Phys. Rev. A}
\bvolume{89},
\bfpage{042311}
(\byear{2014})
\doiurl{10.1103/PhysRevA.89.042311}
\end{barticle}
\endbibitem

\bibitem[\protect\citeauthoryear{Araujo et~al.}{2021}]{Araujo}
\begin{barticle}
\bauthor{\bsnm{Araujo}, \binits{I.F.}},
\bauthor{\bsnm{Park}, \binits{D.K.}},
\bauthor{\bsnm{Petruccione}, \binits{F.}},
\bauthor{\bsnm{Silva}, \binits{A.J.}}:
\batitle{A divide-and-conquer algorithm for quantum state preparation}.
\bjtitle{Scientific Reports}
\bvolume{11}(\bissue{1}),
\bfpage{6329}
(\byear{2021})
\doiurl{10.1038/s41598-021-85474-1}
\end{barticle}
\endbibitem

\bibitem[\protect\citeauthoryear{Murta et~al.}{2023}]{Murta}
\begin{barticle}
\bauthor{\bsnm{Murta}, \binits{B.}},
\bauthor{\bsnm{Cruz}, \binits{P.M.Q.}},
\bauthor{\bsnm{Fern\'andez-Rossier}, \binits{J.}}:
\batitle{Preparing valence-bond-solid states on noisy intermediate-scale
  quantum computers}.
\bjtitle{Phys. Rev. Res.}
\bvolume{5},
\bfpage{013190}
(\byear{2023})
\doiurl{10.1103/PhysRevResearch.5.013190}
\end{barticle}
\endbibitem

\bibitem[\protect\citeauthoryear{Chiribella et~al.}{2013}]{Chiribella}
\begin{barticle}
\bauthor{\bsnm{Chiribella}, \binits{G.}},
\bauthor{\bsnm{D'Ariano}, \binits{G.M.}},
\bauthor{\bsnm{Perinotti}, \binits{P.}},
\bauthor{\bsnm{Valiron}, \binits{B.}}:
\batitle{Quantum computations without definite causal structure}.
\bjtitle{Phys. Rev. A}
\bvolume{88},
\bfpage{022318}
(\byear{2013})
\doiurl{10.1103/PhysRevA.88.022318}
\end{barticle}
\endbibitem

\bibitem[\protect\citeauthoryear{Ara\'ujo et~al.}{2017}]{Baumeler}
\begin{barticle}
\bauthor{\bsnm{Ara\'ujo}, \binits{M.}},
\bauthor{\bsnm{Gu\'erin}, \binits{P.A.}},
\bauthor{\bsnm{Baumeler}, \binits{A.}}:
\batitle{Quantum computation with indefinite causal structures}.
\bjtitle{Phys. Rev. A}
\bvolume{96},
\bfpage{052315}
(\byear{2017})
\doiurl{10.1103/PhysRevA.96.052315}
\end{barticle}
\endbibitem

\bibitem[\protect\citeauthoryear{Castro-Ruiz et~al.}{2018}]{Castro}
\begin{barticle}
\bauthor{\bsnm{Castro-Ruiz}, \binits{E.}},
\bauthor{\bsnm{Giacomini}, \binits{F.}},
\bauthor{\bsnm{Brukner}, \binits{C.}}:
\batitle{Dynamics of quantum causal structures}.
\bjtitle{Phys. Rev. X}
\bvolume{8},
\bfpage{011047}
(\byear{2018})
\doiurl{10.1103/PhysRevX.8.011047}
\end{barticle}
\endbibitem

\bibitem[\protect\citeauthoryear{Preskill}{2018}]{Preskill}
\begin{barticle}
\bauthor{\bsnm{Preskill}, \binits{J.}}:
\batitle{Quantum {C}omputing in the {NISQ} era and beyond}.
\bjtitle{{Quantum}}
\bvolume{2},
\bfpage{79}
(\byear{2018})
\doiurl{10.22331/q-2018-08-06-79}
\end{barticle}
\endbibitem

\bibitem[\protect\citeauthoryear{Jones et~al.}{2019}]{Jones}
\begin{barticle}
\bauthor{\bsnm{Jones}, \binits{T.}},
\bauthor{\bsnm{Endo}, \binits{S.}},
\bauthor{\bsnm{McArdle}, \binits{S.}},
\bauthor{\bsnm{Yuan}, \binits{X.}},
\bauthor{\bsnm{Benjamin}, \binits{S.C.}}:
\batitle{Variational quantum algorithms for discovering hamiltonian spectra}.
\bjtitle{Phys. Rev. A}
\bvolume{99},
\bfpage{062304}
(\byear{2019})
\doiurl{10.1103/PhysRevA.99.062304}
\end{barticle}
\endbibitem

\bibitem[\protect\citeauthoryear{You and Chapman}{2000}]{PhysRevA.62.052302}
\begin{barticle}
\bauthor{\bsnm{You}, \binits{L.}},
\bauthor{\bsnm{Chapman}, \binits{M.S.}}:
\batitle{Quantum entanglement using trapped atomic spins}.
\bjtitle{Phys. Rev. A}
\bvolume{62},
\bfpage{052302}
(\byear{2000})
\doiurl{10.1103/PhysRevA.62.052302}
\end{barticle}
\endbibitem

\bibitem[\protect\citeauthoryear{Barredo et~al.}{2015}]{Barredo2015}
\begin{barticle}
\bauthor{\bsnm{Barredo}, \binits{D.}},
\bauthor{\bsnm{Labuhn}, \binits{H.}},
\bauthor{\bsnm{Ravets}, \binits{S.}},
\bauthor{\bsnm{Lahaye}, \binits{T.}},
\bauthor{\bsnm{Browaeys}, \binits{A.}},
\bauthor{\bsnm{Adams}, \binits{C.S.}}:
\batitle{Coherent excitation transfer in a spin chain of three rydberg atoms}.
\bjtitle{Phys. Rev. Lett.}
\bvolume{114},
\bfpage{113002}
(\byear{2015})
\doiurl{10.1103/PhysRevLett.114.113002}
\end{barticle}
\endbibitem

\bibitem[\protect\citeauthoryear{Labuhn et~al.}{2016}]{Labuhn2016}
\begin{barticle}
\bauthor{\bsnm{Labuhn}, \binits{H.}},
\bauthor{\bsnm{Barredo}, \binits{D.}},
\bauthor{\bsnm{Ravets}, \binits{S.}},
\bauthor{\bsnm{Léséleuc}, \binits{S.}},
\bauthor{\bsnm{Macrì}, \binits{T.}},
\bauthor{\bsnm{Lahaye}, \binits{T.}},
\bauthor{\bsnm{Browaeys}, \binits{A.}}:
\batitle{Tunable two-dimensional arrays of single rydberg atoms for realizing
  quantum ising models}.
\bjtitle{Nature}
\bvolume{534}(\bissue{7609}),
\bfpage{667}--\blpage{670}
(\byear{2016})
\doiurl{10.1038/nature18274}
\end{barticle}
\endbibitem

\bibitem[\protect\citeauthoryear{Barredo et~al.}{2018}]{Barredo2018}
\begin{barticle}
\bauthor{\bsnm{Barredo}, \binits{D.}},
\bauthor{\bsnm{Lienhard}, \binits{V.}},
\bauthor{\bsnm{Léséleuc}, \binits{S.}},
\bauthor{\bsnm{Lahaye}, \binits{T.}},
\bauthor{\bsnm{Browaeys}, \binits{A.}}:
\batitle{Synthetic three-dimensional atomic structures assembled atom by atom}.
\bjtitle{Nature}
\bvolume{561}(\bissue{7721}),
\bfpage{79}--\blpage{82}
(\byear{2018})
\doiurl{10.1038/s41586-018-0450-2}
\end{barticle}
\endbibitem

\bibitem[\protect\citeauthoryear{Ravon et~al.}{2023}]{Ravon2023}
\begin{barticle}
\bauthor{\bsnm{Ravon}, \binits{B.}},
\bauthor{\bsnm{M\'ehaignerie}, \binits{P.}},
\bauthor{\bsnm{Machu}, \binits{Y.}},
\bauthor{\bsnm{Hern\'andez}, \binits{A.D.}},
\bauthor{\bsnm{Favier}, \binits{M.}},
\bauthor{\bsnm{Raimond}, \binits{J.M.}},
\bauthor{\bsnm{Brune}, \binits{M.}},
\bauthor{\bsnm{Sayrin}, \binits{C.}}:
\batitle{Array of individual circular rydberg atoms trapped in optical
  tweezers}.
\bjtitle{Phys. Rev. Lett.}
\bvolume{131},
\bfpage{093401}
(\byear{2023})
\doiurl{10.1103/PhysRevLett.131.093401}
\end{barticle}
\endbibitem

\bibitem[\protect\citeauthoryear{Šibalić et~al.}{2017}]{SIBALIC2017319}
\begin{barticle}
\bauthor{\bsnm{Šibalić}, \binits{N.}},
\bauthor{\bsnm{Pritchard}, \binits{J.D.}},
\bauthor{\bsnm{Adams}, \binits{C.S.}},
\bauthor{\bsnm{Weatherill}, \binits{K.J.}}:
\batitle{Arc: An open-source library for calculating properties of alkali
  rydberg atoms}.
\bjtitle{Computer Physics Communications}
\bvolume{220},
\bfpage{319}--\blpage{331}
(\byear{2017})
\doiurl{10.1016/j.cpc.2017.06.015}
\end{barticle}
\endbibitem

\bibitem[\protect\citeauthoryear{Weber et~al.}{2017}]{Weber_2017}
\begin{barticle}
\bauthor{\bsnm{Weber}, \binits{S.}},
\bauthor{\bsnm{Tresp}, \binits{C.}},
\bauthor{\bsnm{Menke}, \binits{H.}},
\bauthor{\bsnm{Urvoy}, \binits{A.}},
\bauthor{\bsnm{Firstenberg}, \binits{O.}},
\bauthor{\bsnm{Büchler}, \binits{H.P.}},
\bauthor{\bsnm{Hofferberth}, \binits{S.}}:
\batitle{Calculation of rydberg interaction potentials}.
\bjtitle{Journal of Physics B: Atomic, Molecular and Optical Physics}
\bvolume{50}(\bissue{13}),
\bfpage{133001}
(\byear{2017})
\doiurl{10.1088/1361-6455/aa743a}
\end{barticle}
\endbibitem

\bibitem[\protect\citeauthoryear{Ravets et~al.}{2015}]{SR}
\begin{barticle}
\bauthor{\bsnm{Ravets}, \binits{S.}},
\bauthor{\bsnm{Labuhn}, \binits{H.}},
\bauthor{\bsnm{Barredo}, \binits{D.}},
\bauthor{\bsnm{Lahaye}, \binits{T.}},
\bauthor{\bsnm{Browaeys}, \binits{A.}}:
\batitle{Measurement of the angular dependence of the dipole-dipole interaction
  between two individual rydberg atoms at a f\"orster resonance}.
\bjtitle{Phys. Rev. A}
\bvolume{92},
\bfpage{020701}
(\byear{2015})
\doiurl{10.1103/PhysRevA.92.020701}
\end{barticle}
\endbibitem

\end{thebibliography}

\end{document}